\documentclass[12pt,letterpaper]{article}
\pdfoutput=1

\usepackage{amsmath,amssymb,calc, amsthm,bbm, epsfig,psfrag, mathtools}
\usepackage{graphicx, enumerate}
\usepackage[numbers,sort&compress]{natbib}
\usepackage[dvipsnames]{xcolor}
\usepackage{tensor}

\usepackage[utf8]{inputenc} 

\newcommand{\Comment}[1]{{}}
\definecolor{darkblue}{rgb}{0.15,0.35,0.55}
\definecolor{reddish}{rgb}{0.65, 0.2, 0.2}

\usepackage[linktocpage=true]{hyperref}
\hypersetup{
colorlinks=true,
citecolor=darkblue,
linkcolor=reddish,
urlcolor=darkblue,
pdfauthor={},
pdftitle={},
pdfsubject={}
}

\PassOptionsToPackage{linecolor=blue,backgroundcolor=blue!25,bordercolor=blue,textsize=scriptsize}{todonotes}
\usepackage{todonotes}

\makeatletter
\renewcommand\section{\@startsection {section}{1}{\z@}%
                                   {-3.5ex \@plus -1ex \@minus -.2ex}
                                   {2.3ex \@plus.2ex}%
                                   {\normalfont\large\bfseries}}
\renewcommand\subsection{\@startsection{subsection}{2}{\z@}%
                                     {-3.25ex\@plus -1ex \@minus -.2ex}%
                                     {1.5ex \@plus .2ex}%
                                     {\normalfont\bfseries}}
\makeatother

\let\non\nonumber

\newcommand{\bk}{\boldsymbol{k}}

\newcommand{\bL}{\boldsymbol{L}}
\newcommand{\bTheta}{\boldsymbol{\Theta}}
\newcommand{\bomega}{\boldsymbol{\omega}}
\newcommand{\bQ}{\boldsymbol{Q}}
\newcommand{\bE}{\boldsymbol{E}}

\newcommand{\fr}{\mathfrak{r}}
\newcommand{\ffr}{\hat{\mathfrak{r}}}

\def\AdS{{\rm AdS}}

\def\bea#1\eea{\begin{align}#1\end{align}}

\def\bes #1\ees{\begin{split}#1\end{split}}

\newcommand{\be}{\begin{equation}}
\newcommand{\ee}{\end{equation}}

\newcommand{\bma}{\begin{pmatrix}}
\newcommand{\ema}{\end{pmatrix}}

\newcommand{\gh}{\hat{g}}
\newcommand{\Rh}{\hat{R}}
\newcommand{\rh}{\hat{r}}
\newcommand{\kh}{\hat{k}}
\newcommand{\hk}{\hat{k}}
\newcommand{\rt}{\tilde{r}}
\newcommand{\rhot}{\tilde{\rho}}

\newcommand{\Mt}{\widetilde{M}}
\newcommand{\Jt}{\widetilde{J}}

\let\a=\alpha

\let\l=\lambda

\def\e{\epsilon}

\def\l{\lambda}
\def\m{\mu}
\def\n{\nu}

\def\M{{\mathcal M}}

\newcommand{\p}{\partial}

\newcommand{\C}[1]{$(\ref{#1})$}

\def\IZ{\relax\ifmmode\mathchoice
{\hbox{\cmss Z\kern-.4em Z}}{\hbox{\cmss Z\kern-.4em Z}}
{\lower.9pt\hbox{\cmsss Z\kern-.4em Z}} {\lower1.2pt\hbox{\cmsss
Z\kern-.4em Z}}\else{\cmss Z\kern-.4em Z}\fi}
\def\IR{\relax{\rm I\kern-.18em R}}

\def\one{{\hbox{ 1\kern-.8mm l}}}

\newlength{\bredde}
\def\slash#1{\settowidth{\bredde}{$#1$}\ifmmode\,\raisebox{.15ex}{/}
\hspace*{-\bredde} #1\else$\,\raisebox{.15ex}{/}\hspace*{-\bredde}
#1$\fi}

\newsavebox{\zzzbar}
\sbox{\zzzbar}
  {\setlength{\unitlength}{0.9em}
  \begin{picture}(0.6,0.7)
  \thinlines
  \put(0,0){\line(1,0){0.6}}
  \put(0,0.75){\line(1,0){0.575}}
  \multiput(0,0)(0.0125,0.025){30}{\rule{0.3pt}{0.3pt}}
  \multiput(0.2,0)(0.0125,0.025){30}{\rule{0.3pt}{0.3pt}}
  \put(0,0.75){\line(0,-1){0.15}}
  \put(0.015,0.75){\line(0,-1){0.1}}
  \put(0.03,0.75){\line(0,-1){0.075}}
  \put(0.045,0.75){\line(0,-1){0.05}}
  \put(0.05,0.75){\line(0,-1){0.025}}
  \put(0.6,0){\line(0,1){0.15}}
  \put(0.585,0){\line(0,1){0.1}}
  \put(0.57,0){\line(0,1){0.075}}
  \put(0.555,0){\line(0,1){0.05}}
  \put(0.55,0){\line(0,1){0.025}}
  \end{picture}}

\newfont{\goth}{ygoth.tfm scaled 1200}                   

 \numberwithin{equation}{section}

\def\1{{(1)}}
\def\2{{(2)}}
\def\3{{(3)}}

\newcommand{\overbar}[1]{\mkern 1.5mu\overline{\mkern-1.5mu#1\mkern-1.5mu}\mkern 1.5mu}

\def\TT{{T\overbar{T}}}

\def\bk{{\boldsymbol{k}}}

\newcommand{\ul}{\underline}

\begin{document}
\begin{titlepage}

\begin{center}

February 6, 2023
\hfill         \phantom{xxx}  EFI-21-8

\vskip 2 cm {\Large \bf Holography and Irrelevant Operators} 
\vskip 1.25 cm {\bf Chih-Kai Chang$^1$, Christian Ferko$^{2}$ and Savdeep Sethi$^{1}$}\non\\
\vskip 0.2 cm
{\it $^1$ Enrico Fermi Institute \& Kadanoff Center for Theoretical Physics \\ University of Chicago, Chicago, IL 60637, USA}

\vskip 0.2 cm
 {\it $^2$ Center for Quantum Mathematics and Physics (QMAP) \\ Department of Physics \& Astronomy, University of California, Davis, CA 95616, USA}

\vskip 0.2 cm

\end{center}
\vskip 1.5 cm

\begin{abstract}
\baselineskip=18pt\textbf{}

We explore the holographic proposal involving spacetimes with linear dilaton asymptotics in three dimensions from a gravity perspective. The holographic dual shares some properties with a symmetric product conformal field theory deformed by a single-trace analogue of the $\TT$ deformation.
We present solutions of ten-dimensional supergravity which interpolate from BTZ black holes in the interior to either a linear dilaton spacetime near infinity, or to flat space. This allows a precise identification of field theory parameters with gravity parameters. The solutions manifestly exhibit the square root structure that is characteristic of $\TT$-deformed conformal field theories. We compute the mass of the spacetimes using the covariant phase space formalism and find agreement with the square root formula for the case of black holes without spin. We also discuss whether closed string tachyons might play a role when the deformation parameter becomes too large and the vacuum becomes unstable.

\end{abstract}

\end{titlepage}

\tableofcontents

\section{Introduction} \label{intro}

Although holography for asymptotically $\mathrm{AdS}$ spacetimes has been extensively studied and tested in a variety of contexts, holography for spacetimes with non-$\mathrm{AdS}$ asymptotics is poorly understood. This certainly includes the cases of asymptotically flat and asymptotically de Sitter spacetimes.
There are good reasons to suspect that new structures in quantum field theory are needed to define quantum gravity in such spacetimes, should such gravity theories exist. If one assumes a correspondence similar in spirit to the AdS/CFT correspondence then any holographic description is forced to have a high-energy density of states that grows faster than a local quantum field theory in order to match the spectrum of large black holes. If one is in low enough dimension, this obstruction might be avoided but it seems hard to avoid in theories with propagating gravitons.

It seems reasonable then to suspect that a field theory dual to a non-$\mathrm{AdS}$ spacetime, with a potentially rich spectrum of black holes, might be controlled by some structure other than a local quantum field theory at high energies. On the other hand, we have recently learned that some special irrelevant deformations of local quantum field theories possess exactly the property of dramatically modifying the ultraviolet behavior of the theory. It is an exciting prospect that the new structures seen in quantum field theory by turning on controlled irrelevant deformations might play a role in defining quantum gravity on non-$\mathrm{AdS}$ spacetimes.

The most prominent family of controlled irrelevant deformations, which has generated considerable recent excitement, involves operators constructed from bilinears of conserved currents in two-dimensional quantum field theories. The first example in this family is the $\TT$ operator of \cite{Zamolodchikov:2004ce}. This operator, which can be used to deform any $2d$ QFT, is given by the combination $\det(T)$, 
\begin{align}\label{tt_def}
    \TT (x) = \lim_{y \to x} \left( T^{\mu \nu} ( x ) T_{\mu \nu} ( y ) - \tensor{T}{^\mu_\mu} ( x ) \tensor{T}{^\nu_\nu} ( y )  \right) \, , 
\end{align}
where $T_{\mu \nu}$ is the stress tensor of the theory. Although this definition involves a coincident-point limit of local operators, all OPE divergences which arise in this limit are proportional to total derivatives. These total derivatives do not contribute to one-point functions in translationally invariant states. Therefore, up to fairly harmless total derivative ambiguities, one can always define a local irrelevant operator from the stress tensor using (\ref{tt_def}). Since any $2d$ QFT with translation invariance admits a stress-energy tensor, this operator is universal. We will call this the double-trace $\TT$ deformation to distinguish it from a different deformation of interest in this work. 

We define a flow in the space of theories with tangent vector $\int \TT(x)$. The flow parameter $\lambda$ has length dimension $2$. The 
stress tensor must be recomputed at each step along the flow, and the composite operator (\ref{tt_def}) is constructed using this $\lambda$-dependent stress tensor.
Quantizing this theory on a cylinder of radius $L$ gives an energy spectrum which satisfies the inviscid Burgers' equation:
\begin{align}\label{burgers}
        \frac{\partial}{\partial \lambda} E_n ( L, \lambda ) = \frac{1}{2} \frac{\partial}{\partial L} \left( E_n ( L , \lambda )^2 \right) + \frac{P_n(L)^2}{L} \, .
\end{align}
Here $E_n$ are the energies and $P_n$ are the quantized momenta~\cite{Zamolodchikov:2004ce,Smirnov:2016lqw,Cavaglia:2016oda}.

 There is considerable evidence that this deformation defines a theory at the quantum level for flat and even $\AdS_2$ spacetimes~\cite{Brennan:2020dkw}, and leads to a new structure beyond local quantum field theory. For example, the high-energy density of states exhibits a Hagedorn growth which can be inferred from the explicit solution of~\C{burgers} for a $\TT$-deformed $\mathrm{CFT}_2$ on a cylinder of radius $L$, %
  \begin{align} \label{energyformula}
     E_n ( \lambda ) = \frac{L}{2 \lambda} \left( \sqrt{1 + \frac{4 \lambda E_n}{L} + \frac{4 \lambda^2 P_n^2}{L^2}}- 1 \right) \, ,
 \end{align}
where $E_n(\l)$ is the deformed energy, $E_n$ is the undeformed $\mathrm{CFT}_2$ energy and $P_n$ is the momentum of the $n^{\text{th}}$ eigenstate.\footnote{All of our discussion is for the ``good'' sign of the deformation, which corresponds to $\lambda> 0$. For this sign, all of the energies are real for sufficiently small $\lambda$. For the ``bad'' sign $\lambda< 0$, most of the energies are complex.
If one considers sequential flows by irrelevant operators then the bad sign behavior can be cured~\cite{Ferko:2022dpg}.} Many of the basic questions that one might ask about this new structure, like the nature of physical observables, are still unsettled. 

If one instead considers the symmetric product orbifold ${\rm Sym}^N(\mathrm{CFT}_2) =(\mathrm{CFT}_2)^N/S_N$ then we could deform by the single-trace operator
\be\label{singletrace_ttbar}
    D ( x ) = \lim_{y \to x} \sum_{i=1}^N \left[ \left( T_i (x) \right)^{\mu \nu} \left( T_i (y) \right)_{\mu \nu} - \tensor{\left( T_i(x) \right)}{^\mu_\mu} \tensor{\left( T_i (y) \right)}{^\nu_\nu} \right] \, , 
\ee
which exists at the orbifold point~\cite{Giveon:2017nie}, instead of the usual double-trace operator
\begin{align}
    \TT ( x ) = \lim_{y \to x} \left[ \Big( \sum_{i} T_i ( x ) \Big)^{\mu \nu} \Big( \sum_{i} T_i (y) \Big)_{\mu \nu} - \tensor{\Big( \sum_{i} T_i ( x )  \Big)}{^\mu_\mu} \tensor{\Big( \sum_{i} T_i ( y ) \Big)}{^\nu_\nu} \right] \, .
\end{align}
Here $T_i$ denotes the stress tensor for the $i^{\text{th}}$ copy of $\mathrm{CFT}_2$ in the symmetric product. 

There is a very intriguing holographic proposal for defining quantum gravity in spacetimes with linear dilaton asymptotics in three dimensions~\cite{Giveon:2017myj,Giveon:2017nie,Asrat:2017tzd}. Such backgrounds include solutions that interpolate between $\AdS_3$ spacetimes and linear dilaton spacetimes~\cite{Forste:1994wp,Israel:2003ry,Brennan:2020bju}. 
The proposal is that deforming by an operator
like~\C{singletrace_ttbar}, although not precisely this operator,\footnote{The holographic definition of $\mathrm{AdS}_3$ string theory with NS-flux cannot be the symmetric product orbifold, aside possibly from the stringy case of a single NS5-brane~\cite{Eberhardt:2018ouy}. Rather the holographic theory is suspected to correspond to a marginal deformation of the orbifold theory. The existence of a well-defined irrelevant operator analogous to~\C{singletrace_ttbar} away from the orbifold point is in no sense obvious. For our purpose, what is important is that an energy formula like~\C{energyformula} exists for at least some states in the holographic description.}
provides the holographic description of this system~\cite{Giveon:2017myj,Giveon:2017nie,Asrat:2017tzd}. The specific backgrounds studied arise as limits of the gravitational solution describing a collection of type II NS5-branes and fundamental strings. Such a gravitational solution has a holographic description controlled in the ultraviolet by the conjectured little string theory supported on the NS5-branes; for reviews, see \cite{Aharony:1999ks,Kutasov:2001uf}.

This example is especially interesting because an asymptotically linear dilaton (ALD) spacetime is structurally quite different from an asymptotically $\mathrm{AdS}$ spacetime. For instance, $\mathrm{AdS}$ has a timelike boundary but an ALD spacetime has a null component in the boundary. 
%
We might therefore expect holography in the ALD setting to differ significantly from the familiar $\mathrm{AdS}/\mathrm{CFT}$ correspondence. 
Much of the progress in realizing that deformations in the spirit of $D(x)$ play a role in holography has come from worldsheet considerations. In the specific case of the  $M=0$ BTZ background with pure NS-flux, there is a marginal worldsheet deformation that generates the ALD spacetime~\cite{Forste:1994wp,Israel:2003ry}. The effect of this worldsheet deformation on the long string excitation spectrum has been argued to match the way $D(x)$ deforms the energies of the symmetric product orbifold~\cite{Giveon:2017nie}. 
This is compelling evidence that something like a single-trace $\TT$ operator, with a similar effect on the energy spectrum, should exist in the actual holographic $\mathrm{CFT}_2$ which defines $\mathrm{AdS}_3$ with pure NS-flux. 

Our goal in the present work is to provide a complementary view of this holographic proposal: %
rather than using a worldsheet construction, we will work primarily in the target spacetime and perform an analysis using only classical general relativity. We find families of solutions which look like $\AdS_3$ spacetimes in the interior but with either ALD asymptotics \C{spinning_case} or flat space asymptotics \C{final_undecoupled_soln_J_bulk}. What is striking about these solutions, which are parametrized by the mass and spin $(\Mt, \Jt)$ of the interior BTZ black hole, is that the dilaton solution exhibits exactly the square root form seen in \C{energyformula}:
\begin{align}\label{dilaton_form_intro}
    e^{2\Phi} = \frac{r_5^2 \sqrt{1 + 8 k_1\frac{ \Mt  r_5^2}{R^2} + 16 \left(k_1\right)^2 \frac{ \Jt^2 r_5^4}{R^4} }}{\rh_1^2 + k_1 r^2} \, .
\end{align}
The radial coordinate in \C{dilaton_form_intro} is $r$ while the parameter $k_1$ corresponds to $\l$ of \C{energyformula} via the relation:
\begin{align}\label{lambda_intro}
    \l = \frac{  (\a')^2 |m_5|}{2R^2} k_1\, .
\end{align}
The constants $r_5$ and $\rh_1$ in (\ref{dilaton_form_intro}) are fixed length scales determined by the number of NS5-branes, $m_5$, and the number of fundamental strings, $m_1$. The string length is $\ell_s = \sqrt{\alpha'}$ and $R$ is the size of an asymptotic circle which is finite in string-frame and characteristic of the ALD spacetime.

It is worth noting that the effect of the $k_1$ deformation is detectable even in the deep interior where $r$ is very small. For example, the $\AdS_3$ length scale is $k_1$-deformed in a mass and spin-dependent way. Alternately, the value of the dilaton at $r=0$ in \C{dilaton_form_intro} is now mass and spin-dependent while it is simply constant in the case of $\AdS_3$. Said differently: the gravity solution `knows' about the irrelevant deformation of the holographic CFT even in regions of spacetime normally associated to the deep infrared of the CFT.
%

Our conventions can be found in section~\ref{sec:review} along with a review of the $\Mt=0$ brane solutions.  In section \ref{sec:generalized_solutions} we summarize the final form of the various families of solutions with mass and spin. A more detailed 
analysis leading to these solutions can be found in Appendix \ref{app:derivation_details}. In section \ref{sec:mass}, we use the covariant phase space formalism to compute the mass of the ALD solutions. For the case without spin, we find a square root form for the mass of the spacetime \C{sqrt_charge_soln} in accord with our expectations for a $\TT$-like deformation of the holographic description. The case with spin is more subtle because of issues with charge integrability which we discuss in section \ref{sec:non-rotating}. Lastly, in section \ref{tachyon} we briefly explore whether a closed string tachyon might play a role in understanding the fate of the vacuum state when the deformation parameter becomes too large.

Some generalizations of the holography discussed here for the pure NS background to cases with $(p,q)$ $5$-branes, which really require the kind of gravity approach we have employed, will appear elsewhere.

\vskip 0.1in
\noindent {\bf Note Added:} The results in this paper have been in a state of partial completion for quite a long time. During that period two interesting papers appeared with related results from different approaches \cite{Apolo:2019zai,Apolo:2021wcn}. The long delay can, in part, be attributed to  the arrival of a new baby during the COVID lockdown. After this paper appeared, we were informed of other potentially related work~\cite{Chakraborty:2020swe}.

\section{Type IIB Supergravity Brane Solutions}\label{sec:review}

We first want to find general solutions of type II supergravity with NS5-brane and fundamental string charge that asymptote to flat space. There will be two decoupling limits that play a role in our discussion and we will want to write the solutions in convenient coordinates for exhibiting these limits. 

As a starting point, in this section we review brane solutions of type IIB supergravity, which fully decouple to the Poincar\'e patch of the $M=0$ BTZ background. In section~\ref{sec:generalized_solutions}, we will find the general class of solutions with non-zero mass and spin. Because the solutions involve a torus, we can dualize between type IIA or type IIB string theory; for concreteness, we will discuss solutions of the type IIB equations of motion.

\subsection{Constraints from Flux Quantization} \label{flux}

To fix conventions, let us begin with the bosonic action for ten-dimensional type IIB supergravity in string-frame:
\begin{align}\label{IIB_action}
    S_{{\rm IIB}} = {1\over (2\pi)^7 {\alpha'}^4}\int d^{10}x \, \sqrt{- \gh} \Bigg\{ & e^{-2\Phi} \left( \Rh + 4 (\p \Phi)^2 - {1\over 12} |H_3|^2\right) - {1\over 2} |F_1|^2 \cr & - {1\over 12}|F_3 - C_0 H_3|^2 - \frac{1}{4 \cdot 5!} | \widetilde{F}_5 |^2 \Bigg\} ,
\end{align}
where $\boldsymbol{\widetilde{F}}_5 = d \boldsymbol{C}_4 - {1\over 2} \boldsymbol{C}_2 \wedge \boldsymbol{H}_3 + {1\over 2} \boldsymbol{B}_2 \wedge \boldsymbol{F}_3$.\footnote{We use boldface to denote differential forms to conform with the notation often found in the gravity literature and used in section \ref{sec:mass}. } We define the gravitational constant 
\begin{align}
    \kappa_{10}^2 := {1\over 2} (2\pi)^7 {\alpha'}^4 e^{2\Phi_0} \, ,
\end{align}
in terms of the string length, $\ell_s=\sqrt{\alpha'}$, and the asymptotic value of the string coupling $g_s = e^{\Phi_0}$. The hatted variables $\gh$ and $\Rh$ refer to string-frame quantities. 

The analysis of $\TT$-like deformations and holography in general flux backgrounds -- when the Ramond fluxes $C_0, \boldsymbol{C}_2$, $\boldsymbol{C}_4$ may be nonzero -- will be explored in future work. For our present purposes, we will restrict to the case of pure NS fluxes. In this case, it is consistent to set $C_0 = \boldsymbol{C}_2 = \boldsymbol{C}_4 = 0$.

We will consider solutions of the form $\mathcal{M}_3 \times S^3 \times T^4$. The $T^4$ factor is a spectator which could also be replaced by K3 without changing the analysis. Although not visible in supergravity, in string theory flux quantization through the $S^3$ requires that 
\begin{align}\label{H3_quantization}
{1\over 4\pi^2 \alpha'} \int_{S^3} \boldsymbol{H}_3 = m_5 \, , 
\end{align}
where $m_5$ is an integer corresponding to the NS$5$-brane charge.

There is another quantization condition on the dual field strength. If we define
\begin{align}\label{H7_quantization}
    \boldsymbol{H}_7 = e^{-2\Phi} \ast \boldsymbol{H}_3 \, , 
\end{align}
then this form satisfies the following quantization condition for any $7$-cycle $\Sigma_7$, 
\begin{align}\label{dualquantization}
  {1\over (2\pi)^6 (\alpha')^3}  \int_{\Sigma_7} \boldsymbol{H}_7 = m_1 \in \mathbb{Z} \, .
\end{align}

\subsection{Review of NS5-Brane and F-String Bound State}

There is a well-known class of solutions in type IIB supergravity which satisfy the flux quantization conditions outlined in the previous subsection and which form the prototype for the generalized solutions that we will consider in section \ref{sec:generalized_solutions}. These much-studied solutions are interpreted as bound states of $m_1$ fundamental strings and $m_5$ NS5 branes, where $m_1, m_5$ are the integers appearing in the quantization conditions (\ref{dualquantization}) and (\ref{H3_quantization}), respectively.

One way of presenting these solutions \cite{Tseytlin:1996as}, which is nicely reviewed in \cite{Kluson:2016dca}, is
\begin{align}\label{kluson_soln}
 &   e^{-2 \Phi} = \frac{1}{g_s^2} \frac{f_1}{f_5} \, , \qquad B_{05} = \frac{1}{f_1} - 1 \, , \qquad H_{mnp} = \tensor{\epsilon}{_m_n_p^q}  \, \partial_q (\log f_5) 
 \nonumber \\
  &  \widehat{ds}^2 = \frac{-dt^2 + dx_5^2}{f_1} + f_5 \left( dx_1^2 + \cdots + dx_4^2 \right) + \left( dx_6^2 + \cdots dx_9^2 \right) \, , 
\end{align}
where $m, n, p, q = 1,\ldots, 4 $, $\e_{mnpq}$ is the volume form for $f_5 \sum_{i=1}^4 dx_i^2$ and
\begin{align}\label{kluson_f1_f5_def}
    f_1 = 1 + \frac{ r_1^2}{r^2} = 1 + \frac{16 \pi^4 g_s^2 \alpha^{\prime 3} m_1}{V_4 r^2} \, , \qquad f_5 = 1 + \frac{r_5^2}{r^2} = 1 + \frac{\alpha' m_5}{r^2} \, .
\end{align}
In these expressions, $r$ represents a radial coordinate in the four-dimensional space transverse to the NS5-branes parameterized by $x_1, x_2, x_3, x_4$, while the coordinates $x^6, x^7, x^8, x^9$ parametrize a torus with volume $V_4$. The asymptotic region corresponds to large $r$. The constant $g_s$ specifies the asymptotic value of the string coupling. We have a choice about how to treat $x_5$. To patch to $\AdS_3$ in the interior, we will periodically identify
\begin{align}\label{periodicity_x5}
    x_5 \sim x_5 + 2\pi R \, .
\end{align} 
Note that the flux $H_3$ has two separate contributions, one arising from the explicit $H_{mnp}$ components and one arising from $\boldsymbol{H} = d \boldsymbol{B}$ with $\boldsymbol{B} = B_{05} \, dx^0 \wedge dx^5$ and $B_{05}$ as indicated in (\ref{kluson_soln}).

The solution (\ref{kluson_soln}) contains at least two qualitatively different asymptotic regions. For $r \gg r_1, r_5$, the solution approaches flat $10$-dimensional Minkowski space. On the other hand, in the near-horizon limit $r \ll r_1, r_5$, the sphere and torus become spectators and we recover an effective three-dimensional gravity solution which is diffeomorphic to a Poincar\'e patch of $\mathrm{AdS}_3$.

We will refer to the full spacetime with the asymptotically flat region retained as the undecoupled or totally undecoupled solution, and the near-horizon solution in the deep bulk as the decoupled or totally decoupled solution. The process of taking the full decoupling limit 
may be thought of as simultaneously sending the asymptotic string coupling $g_s\rightarrow 0$ and sending $\alpha' \to 0$, along with a re-scaling of coordinates. Alternatively, one may think of this decoupling limit as dropping the $1$'s in the harmonic functions $f_1$ and $f_5$ so that $f_i = 1 + \frac{r_i^2}{r^2} \, \to \, f_i = \frac{r_i^2}{r^2}$.

\subsubsection*{\ul{\it Partial Decoupling}}

There is also an intermediate regime in this spacetime which looks neither like $\mathrm{AdS}_3$ in the Poincar\'e patch nor like flat space. To take this limit, define $r = g_s \rh$ and take $g_s\to 0$ giving, 
\begin{align}\begin{split} 
     &   e^{-2 \Phi} =  \frac{\rh^2}{r_5^2} \left( 1 + \frac{\rh_1^2}{\rh^2} \right) \, , \quad\qquad \widehat{ds}^2 = \frac{-dt^2 + dx_5^2}{f_1} + \frac{r_5^2}{\rh^2} d\rh^2 +  r_5^2 d\Omega_3^2+ ds^2_{T^4} \, , \label{partialasymptotics}\\
     & f_1 = 1 + \frac{\rh_1^2}{\rh^2} = 1 + \frac{16 \pi^4 \alpha^{\prime 3} m_1}{V_4 \, \rh^2}, \qquad r_1 = g_s \rh_1 \, .
\end{split}\end{align}
This amounts to replacing $f_5 = 1 + \frac{r_5^2}{r^2}$ by $f_5 = \frac{r_5^2}{r^2}$, but retaining the full function $f_1$ without dropping the $1$ there, since $\frac{r_1^2}{r^2}$ is not necessarily big.

\subsubsection*{\ul{\it Full Decoupling}}

The fully decoupled solution is found by making $r$ very small and dropping the  $1$ in $f_1$~\cite{Maldacena:1997re}. To put the resulting metric in a conventional $\AdS_3$ form, we define dimensionless variables 
\begin{align}
    \rt^2 = \rh^2 \frac{R^2}{r_5^2\, \rh_1^2}, \qquad \tilde\varphi = \frac{x_5}{R} \sim \tilde\varphi + 2\pi, \qquad {\tilde t} = \frac{t}{R}. 
\end{align}
In these variables the metric takes the form
\begin{align}\begin{split}\label{full_decoupling}
    \widehat{ds}^2 &= \ell^2 \left( \rt^2 \left( -d{\tilde t}^2 + d\tilde\varphi^2 \right) + \frac{d\rt^2}{\rt^2} + d\Omega_3^2\right) + ds^2_{T^4} \, , \\
    e^{-2 \Phi} &= \frac{\rt^2 r_1^2}{R^2} + \frac{\rh_1^2}{\ell^2} \, \xrightarrow{\text{decoupling}} \, \frac{\rh_1^2}{\ell^2} \, ,
\end{split}\end{align}
where $\ell = r_5$ is the $\AdS_3$ length scale. We note that the size, $R$, of the $x_5$ circle at infinity is absorbed into a redefinition of time and therefore energy. This will play a role later when we compare energies in the partially decoupled theory to energies in the $\AdS_3$ region. The last comments on the fully decoupled case concern the value of the gravitational coupling in the ${\rm AdS}_3$ theory which is given by,
\begin{align}\label{kappa3_defn}
   \left( \kappa_3 \right)^2 = 8\pi G_3= \frac{2\pi \ell_s}{|m_1| \sqrt{|m_5|}},
\end{align}
and the central charge of the dual holographic CFT which is given by 
\begin{align}\label{central_charge}
    c = \frac{3 \ell}{2 G_3} = 6 |m_1 m_5| \, .
\end{align}
We can identify the radius $L$ of the holographic CFT by noting that the mass of global $\AdS_3$ is $-\frac{1}{8G_3}$. This corresponds to the ground state of the CFT with energy $-\frac{c}{12L}$ so we identify:
\begin{align}
    L = \ell = \sqrt{|m_5|} \ell_s \, .
\end{align}
This is independent of the radius $R$ of the $x_5$ circle which appears in \C{periodicity_x5}. 

As we discussed in the introduction, gravity solutions of the partially-decoupled form have been argued to have field theory duals which are obtained by deforming a conformal field theory by an irrelevant operator related to $\TT$~\cite{Giveon:2017myj,Giveon:2017nie,Asrat:2017tzd}. These solutions have also been shown to arise from TsT transformations~\cite{Araujo:2018rho,Apolo:2019zai,Apolo:2021wcn}.
In the following section we will generalize these well-known solutions for the $M=0$ BTZ case without spin to metrics which interpolate from a general spinning BTZ black hole in the interior to either an asymptotically linear dilaton spacetime or to an asymptotically flat spacetime.

\section{Generalized NS5-Brane and F1-String Solutions}\label{sec:generalized_solutions}

In this section, we will present a class of supergravity solutions with the desired quantized charges, symmetries and asymptotic behavior. These solutions can be interpreted as bound states of a system of F1 strings and NS5-branes that interpolate between an $\mathrm{AdS}_3$ black hole in the deep interior with some mass and spin to either a linear dilaton solution in the asymptotic region, or to asymptotically flat six dimensions at large distances. We will separately present those two cases because the latter is a more complicated class of solutions.

The full spacetime equations of motion are quite difficult to solve in generality without some physically reasonable metric ansatz as an input. Our ansatz is spelled out in Appendix \ref{app:derivation_details} along with the calculations of the metric and dilaton functions, and the details of the fairly involved asymptotic matching needed to identify the physical parameters of the solution. The aim of this section is to present clearly the assumptions going into the analysis and the final solutions.  There are other classes of solutions in Appendix \ref{app:derivation_details}, which are even more surprising: they appear to interpolate from an $\AdS_3$ spacetime in the interior to an asymptotic $3$-dimensional spacetime with positive curvature. Whether those solutions are fully sensible will be explored elsewhere.

\subsection{Characterizing the General Solution} \label{subsec:description_general_solution}

We will look for solutions to the equations of motion of type IIB supergravity, written using the same coordinates as \C{kluson_soln}, which satisfy the following assumptions:

\begin{enumerate}
    \item\label{assumption_asymptotics} Asymptotics. We assume that near infinity the family of solutions approach either the linear dilaton asymptotic form seen in \C{partialasymptotics}, or the flat space asymptotic form seen in \C{kluson_soln}. We will insist that the isometries of $T^4$, $S^3$ and $S^1$ are preserved by the solutions. This collapses the unknown functions in the metric, fluxes and dilaton to functions of $r$ and possibly time $t$. We assume these coordinates can be extended to global coordinates for our spacetime soution. Lastly, we assume the metric, dilaton, and NS-flux  admit an expansion in powers of $\frac{1}{r}$ near $r \to \infty$.

    Near $r=0$, we insist the solution takes the form $\mathcal{M}_3 \times S^3 \times T^4$ with the dilaton $\Phi$ 
    tending to a constant. We need this form if we are to interpret the holographic theory as some kind of deformed CFT.

    \item\label{assumption_symmetries} Symmetries. We assume that the solution is stationary and therefore only depends on the radial coordinate $r$. 
    Because the spacetime is assumed to be stationary but not static, the timelike Killing vector $\partial_t$ need not be irrotational. However, we will constrain the allowed rotation to occur only in the $x_5$ direction. Operationally, this means that a non-vanishing metric component $g_{t x_5}$ is allowed -- corresponding to a spinning BTZ black hole at small $r$ -- but no other off-diagonal metric components are permitted.

    \item\label{assumption_fluxes} Fluxes. We assume that the Ramond potentials $\boldsymbol{C}_p$ are all vanishing and that the NS flux $\boldsymbol{H}_3 = d \boldsymbol{B}_2$ only threads the three-dimensional submanifolds $S^3$ and $\mathcal{M}_3$. So $\boldsymbol{H}_3$ is a sum of the volume forms $\boldsymbol{\epsilon}_{S^3}$ and $\boldsymbol{\epsilon}_{\mathcal{M}_3}$ on $S^3$ and $\mathcal{M}_3$, respectively, with coefficients that can in principle depend on $r$. Finally, we impose the usual flux quantization conditions~\C{H3_quantization} and~\C{dualquantization} on $H_3$ required by string theory.

\end{enumerate}

\subsection{Solutions with ALD Asymptotics}
\subsubsection*{\ul{\it The $\Jt=0$ case}}
From the analysis presented in Appendix \ref{app:derivation_details}, we find the following solutions in string-frame for the case where the black hole has no spin: 
\begin{align}\label{final_string_frame_soln_bulk}
   &  ds^2 = - \frac{1 - \frac{ \alpha_e^2}{r^2}}{k_1 + \frac{\rh_1^2}{r^2}} \, dt^2 + \frac{1}{k_1 + \frac{\rh_1^2}{r^2}} \, dx_5^2 + \frac{ r_5^2}{ r^2- \alpha_e^2}  \, dr^2 + r_5^2 \, d \Omega_3^2 + ds^2_{T^4} \, , \nonumber \\
&  e^{2\Phi} = \frac{r_5^2 \sqrt{1 + 8 k_1\frac{ \Mt  r_5^2}{R^2} }}{\rh_1^2 + k_1 r^2} \, , \qquad \alpha_e^2 = 8 \Mt \frac{\rh_1^2 r_5^2}{R^2}\, , \quad r_5^2 = \alpha' m_5\, , \quad \rh_1^2=\frac{16 \pi^4 \alpha^{\prime 3} m_1}{V_4}\, .
\end{align}
These solutions depend on a dimensionless parameter $k_1$. The fully decoupled pure $\AdS_3$ solutions correspond to $k_1=0$. We will momentarily identify $k_1$ with the deformation parameter $\lambda$ of a $\TT$-deformed theory. It is important to point out that the $\Mt=0$ solutions, which match \C{partialasymptotics}, are sensitive to whether $k_1$ is zero or non-zero. However, the precise value of $k_1$ is not important and can be absorbed in a rescaling of coordinates. This reflects the fact that $E=0$ states remain zero energy under a $\TT$ deformation: namely, zero energy does not flow. Any case with $\Mt \neq 0$, however, is sensitive to the value of $k_1$.

Note that the $x_5$ circle  has periodicity $\frac{2\pi R}{\sqrt{k_1}}$ at $r=\infty$ in string-frame. In Einstein frame, the $x_5$ circle has a proper size that goes to infinity as $r\rightarrow \infty$. The dimensionless mass $\Mt$ is related to the dimensionful mass of the BTZ black hole near $r=0$ via $\Mt = M G_3$, where $G_3$ is specified in \C{kappa3_defn}.

On first glance, it might appear that the solutions \C{final_string_frame_soln_bulk} have the wrong large $r$ asymptotics in Einstein-frame because the dilaton is mass-dependent. However a careful redefinition of the radial coordinate in the large $r$ region, discussed around \C{defrho}, shows that these solutions have the correct asymptotic behavior. Also notice that the value of the dilaton at $r=0$ is now mass-dependent and differs from the pure $\AdS_3$ case where $k_1=0$. As a consequence, the $\AdS_3$ length scale is also mass-dependent in Einstein frame. 

Global $\AdS_3$ has mass $\Mt = - \frac{1}{8}$. This corresponds to the ground state of the holographic CFT which should have dimensionless energy $\widetilde{E} = E L = \Mt \frac{L}{G_3}$. Notice that the dilaton solution is only real if the condition, 
\begin{align}\label{gravity_condition}
    k_1 < \frac{R^2}{r_5^2} \, ,
\end{align}
is satisfied. It is important to note that this condition is independent of $m_1$ but depends on the value of $m_5$ via $r_5^2$. As we increase $m_5$, the maximum value of $k_1$ decreases. It is this asymmetry between $m_1$ and $m_5$ that singles out the single-trace $\TT$ deformation rather than the conventional double-trace deformation. In the latter case, we should have seen the maximum value of $k_1$ reduce as we increase the total central charge \C{central_charge} which depends on both $m_1$ and $m_5$. We can then compare this bound with (\ref{energyformula}) which states that 
\begin{align}
    \lambda < \frac{3L^2}{c_{\rm block}} = \frac{3L^2}{6|m_5|}\, ,
\end{align}
where $c_{\rm block} = 6|m_5|$ is the central charge of a block that would appear if the holographic dual were actually a symmetric product~\cite{Giveon:2017nie}. Via this comparison we can identify, 
\begin{align} \label{k1}
    \l = \frac{  (\a')^2 |m_5|}{2R^2} k_1\, , \quad \qquad {\tilde \l} = \frac{\lambda}{L^2} = \frac{\a'}{2R^2} k_1\, ,
\end{align}
where we have also defined the natural dimensionless $\TT$-deformation parameter $\tilde{\l}$. This makes clear that the $\TT$ deformation is controlled by the asymptotic size of the $x_5$ circle in string frame. 
We will see further evidence for this identification in the spinning case. It is striking that the square root solution of the $\TT$ flow equation for a CFT emerges directly from gravity in the structure of the dilaton solution. 

\subsubsection*{\ul{\it The $\Jt \neq 0$ case}}

The solutions for spinning black holes are similar, 
\begin{align}\label{spinning_case}
   &  ds^2 = - \frac{1 - \frac{ \alpha_e^2}{r^2} + \frac{\alpha_j^4}{r^4}}{k_1 + \frac{\rh_1^2}{r^2}} \, dt^2 + \frac{1}{k_1 + \frac{\rh_1^2}{r^2}} \, \left( dx_5 - \frac{\alpha_j^2}{r^2} dt\right)^2 + \frac{r_5^2}{r^2 - \alpha_e^2 + \frac{\alpha_j^4}{r^2}}  \, dr^2 + r_5^2 \, d \Omega_3^2 + ds^2_{T^4} \, , \nonumber \\
&  e^{2\Phi} = \frac{r_5^2 \sqrt{1 + 8 k_1\frac{ \Mt  r_5^2}{R^2} + 16 \left(k_1\right)^2 \frac{ \Jt^2 r_5^4}{R^4} }}{\rh_1^2 + k_1 r^2} \, , \qquad  \alpha_e^2 = 8 \Mt \frac{\rh_1^2 r_5^2}{R^2}\, , \qquad \a_j^2 = 4 \Jt \, \frac{ \rh_1^2 r_5^2}{R^2} .
\end{align}
Here $\Jt= \frac{J G_3}{\ell}$ is a dimensionless spin, but this is not the natural quantity to study from the perspective of the dual CFT. The $\mathrm{AdS}/\mathrm{CFT}$ correspondence maps the dimensionless parameter $J = \widetilde{J} \frac{\ell}{G_3}$ to the dimensionless momentum quantum number $\widetilde{P} = P L$ of the holographic CFT. With this identification, the form of the dilaton solution again beautifully matches the structure of the $\TT$-deformed energy formula~\C{energyformula}, 
\begin{align}
    \sqrt{1 + 8 k_1\frac{ \Mt  r_5^2}{R^2} + 16 \left(k_1\right)^2 \frac{ \Jt^2 r_5^4}{R^4} } \quad \longrightarrow \quad\sqrt{ 1 + 4 \widetilde{\lambda} \widetilde{E} + 4 \widetilde{\lambda}^2 \widetilde{P}^2 } \, ,
\end{align}
where we identify
\begin{align}
    \widetilde{P} = \frac{J}{m_1} \, , \qquad \widetilde{E} = 4 m_5 \widetilde{M} = \frac{1}{m_1} \left( \widetilde{M} \cdot \frac{\ell}{G_3} \right) \, .
\end{align}
Here $\widetilde{E}$ and $ \widetilde{P}$ are the energy and momentum, respectively, of a single block of the symmetric product.

\subsection{Solutions with Flat Space Asymptotics}

Lastly we turn to the fully undecoupled solutions that asymptote to a flat $6$-dimensional spacetime. In string-frame, the metric and dilaton take the form:
\begin{align}\label{final_undecoupled_soln_J_bulk}
   &  ds^2 = - \frac{1 - \frac{ \alpha_e^2}{r^2} + \frac{\alpha_j^4}{r^4}}{k_1 + \frac{\rh_1^2}{r^2}} \, dt^2 + \frac{1}{k_1 + \frac{\rh_1^2}{r^2}} \, \left( dx_5 - \frac{\alpha_j^2}{r^2} dt\right)^2  \nonumber \\
   &\quad  + \frac{k_5 r^2 - \frac{k_5 \alpha_e^2}{2} + \sqrt{ r_5^4 - k_5^2 \alpha_j^4 + \frac{1}{4} k_5^2 \alpha_e^4}}{r^4 - r^2 \alpha_e^2 + \alpha_j^4}  \, r^2  \, dr^2  + r^2 \left( k_5 + \frac{\alpha_5^2}{r^2} \right) \, d \Omega_3^2 + ds^2_{T^4} \, , \nonumber \\
    &  e^{2 \Phi} = \frac{g_s^2 + \frac{\gamma \alpha_5^2}{k_1 r^2}}{1 + \frac{\rh_1^2}{k_1 r^2}}  \, , \qquad  \alpha_e^2 = 8 \Mt \frac{\rh_1^2 r_5^2}{R^2}\, , \qquad \a_j^2 = 4 \Jt \, \frac{ \rh_1^2 r_5^2}{R^2} \, , \qquad k_5 = \frac{k_1 g_s^2}{\gamma} \, ,  \nonumber \\
    &\alpha_5^2 = \frac{1}{2} \left( \sqrt{ k_5^2 \alpha_e^4 + 4 \left( r_5^4  - k_5^2 \alpha_j^4 \right)} - k_5 \alpha_e^2 \right)  \, , \qquad \gamma= \sqrt{ 1 + \frac{k_1 \alpha_e^2}{\rh_1^2} + \frac{k_1^2 \alpha_j^4}{\rh_1^4} } \, .
\end{align}
This final form for the metric depends on the dimensionless parameters $(g_s, k_1)$ where $g_s$ is the $r=\infty$ value of the string coupling. The solution also depends on the flux quantum numbers via $(\rh_1, r_5)$ defined in \C{final_string_frame_soln_bulk} and the black hole mass and spin $(\Mt, \Jt)$. We note that the $3$-sphere now grows as $r$ becomes large which is why the asymptotic theory is $6$-dimensional. 

The square root structure of the solution to the $\TT$ flow equation is encoded in $\gamma$ which now appears both in the metric via $k_5$ and the dilaton. Notice that the size of the $S^3$ at $r=0$ is also deformed from the pure $\AdS_3$ case by an amount that depends on the mass and spin of the interior black hole. Lastly we note that the parameter $k_1$ still controls the size of the $x_5$ circle near infinity.   

\section{Mass Calculation} \label{sec:mass}

The supergravity solutions derived in the previous section appear to have something to do with a $\TT$-deformed theory. 
We made a tentative identification of the deformation parameter with $k_1$ of the spacetime solution based on the appearance of the square root in the dilaton solution of~\C{spinning_case}. This identification assumes that the holographic CFT is of symmetric product type and that the deformation is purely a single-trace $\TT$ deformation. The only cases where we have formulae for the deformed energies are the single and double-trace $\TT$ deformations, and sequential flows by such irrelevant operators~\cite{Ferko:2022dpg}. However, the undeformed holographic CFT is not a symmetric product though it is believed to be connected to a symmetric product by a marginal deformation. Regardless, we can provide further evidence for a correspondence with some kind of single-trace $\TT$-like deformation of that undeformed theory by directly computing the mass of the solutions  \C{final_string_frame_soln_bulk}.

The definition of mass in general relativity is not unique. There are a variety of proposals that depend on the asymptotics of the background. A way to uniformly define the mass is by using the covariant phase space formalism~\cite{Iyer:1994ys}. This will provide an independent way of confirming the parameter identification we made in section \ref{sec:generalized_solutions}.

\subsection{Surface Charges in Gravitational Theories}\label{sec:charge_review}

The covariant phase space formalism allows us to define conserved surface charges in generally covariant theories. In this section we will review this formalism before using it to compute the charges associated with the solutions in section \ref{sec:generalized_solutions}. Our discussion of this formalism will be very brief and many facts will be stated without justification; for a more comprehensive treatment, we refer the reader to a review such as \cite{Compere:2018aar}.

First consider a theory described by a general Lagrangian density $\boldsymbol{L} ( \Phi^i )$ in a theory with $n$ spacetime dimensions, where $\Phi^i$ represents an arbitrary collection of fields. Following common conventions in the gravity literature we denote differential forms in spacetime by boldface symbols, so $\boldsymbol{L} = L \, d^n x$ is the top form associated with the usual Lagrangian density $L$. In pure Einstein gravity, for instance, the only field $\Phi^i$ is the metric $g_{\mu \nu}$ and $L = \frac{1}{16 \pi G} \sqrt{-g} R$. Upon variation of the fields $\Phi^i$, one sees
\begin{align}
    \delta \bL = \frac{\delta \bL}{\delta \Phi^i} \, \delta \Phi^i - d \bTheta [ \delta \Phi^i ; \Phi^i ] \, .
\end{align}
The form $\bTheta [ \delta \Phi^i ; \Phi^i ]$ is called the presymplectic potential and depends both on the fields $\Phi^i$ and their variations $\delta \Phi^i$. One can think of $\bTheta$ as simultaneously being an $(n-1)$-form in spacetime and a $1$-form in the space of field variations.  In what follows, we will suppress the index $i$ when writing the functional dependence on fields and simply write $\bTheta [ \delta \Phi ; \Phi ]$. To define conserved charges, it is convenient to introduce a related object $\bomega$ called the presymplectic form via
\begin{align}
    \bomega [ \delta_2 \Phi , \delta_1 \Phi ; \Phi ] = \delta_2 \bTheta [ \delta_1 \Phi ; \Phi ] - \delta_1 \bTheta [ \delta_2 \Phi ; \Phi ] \, .
\end{align}
Here we write the dependence of $\bomega$ on both variations $\delta_{1,2} \Phi$ to emphasize that $\bTheta$ itself depends on one variation $\delta_1 \Phi$ of the fields, and $\bomega$ is defined by again performing a second variation $\delta_2$ of the fields and finding how $\bTheta$ varies (with appropriate antisymmetrization).  From this perspective, the symbol $\delta$ acts as the exterior derivative in field space so that $\bomega$ is an $(n-1)$-form in spacetime and a $2$-form in the space of field variations.

We will be especially interested in field variations $\delta_\xi \Phi$ that are associated with an infinitesimal diffeomorphism generated by $x^\mu \to x^\mu + \xi^\mu$. For instance, under such a transformation any tensor field $T_{\mu_1 \cdots \mu_p}$ transforms via the Lie derivative $\mathcal{L}_\xi$ as
\begin{align}
    \delta_\xi T_{\mu_1 \cdots \mu_p} = \mathcal{L}_\xi T_{\mu_1 \cdots \mu_p} \, .
\end{align}
We will consider the case where the field configuration $\Phi$ satisfies the equations of motion for the theory, the first variation $\delta_1 \Phi = \delta \Phi$ solves the linearized equations of motion about the solution $\Phi$, and the second variation $\delta_2 \Phi = \delta_\xi \Phi$ is generated by a diffeomorphism of this form. In this case, one can show that $\bomega [ \delta_\xi \Phi , \delta \Phi ; \Phi ]$ is exact, so that
\begin{align}
    \bomega [ \delta_\xi \Phi , \delta \Phi ; \Phi ] = d \bk_\xi [ \delta \Phi ; \Phi ] \, .
\end{align}
This differential form $\bk_\xi$ is an $(n-2)$-form in spacetime and a $1$-form in the space of field variations. Note that we will also identify the $(n-2)$-form $\bk_\xi$ with its Hodge dual, an anti-symmetric $2$-tensor with components $k^{\mu \nu}$, which we write without boldface type. This $\bk_\xi$ can also be expressed as
\begin{align}
    \bk_\xi [ \delta \Phi ; \Phi ] = - \delta \bQ_\xi [ \delta \Phi ; \Phi ] + i_\xi \bTheta [ \delta \Phi ; \Phi ] \, , 
\end{align}
up to the ambiguity of adding a total (spacetime) derivative term to $\bk_\xi$. This ambiguity will not be relevant for us, since we are primarily interested in the integral
\begin{align}\label{change_in_charge}
    \delta Q_\xi = \oint_S \bk_\xi [ \delta \Phi ; \Phi ] \, ,
\end{align}
where $S$ is a closed codimension-$2$ surface. Typically we will think of a spacetime with a radial direction $r$ and let the surface $S$ be a sphere at a fixed (large) value of $r$ and at a fixed time $t$. For this case, the total derivative ambiguity of $\bk_\xi$ will not contribute to the integral in (\ref{change_in_charge}).

Finally, we will specialize to the case where $\xi$ is an exact Killing vector of the solution. In this case, the expression $\delta Q_\xi$ in (\ref{change_in_charge}) can always be integrated in order to define a conserved charge $Q_\xi$. Operationally, this procedure always involves first computing the \emph{change} in the charge $Q_\xi$ associated with a variation $\delta \Phi$ around a particular solution $\Phi$. In examples of interest, we usually think of a one-parameter family of solutions $\Phi ( \alpha )$ which solve the equations of motion for a continuous range of values of $\alpha$. In this case, we can consider the variation
\begin{align}\label{variation_parameter_g}
    \delta_\alpha \Phi = \frac{\partial \Phi ( \alpha )}{\partial \alpha} \, .
\end{align}
Since $\Phi ( \alpha )$ is a solution for any $\alpha$, the variation (\ref{variation_parameter_g}) always satisfies the linearized equations of motion. We can then compute
\begin{align}\label{change_in_charge_alpha}
    \frac{d Q_\xi}{d \alpha} = \oint_S \bk_\xi [ \delta_\alpha \Phi ; \Phi ] \, ,
\end{align}
and integrate the resulting expression with respect to $\alpha$ (using a suitable initial condition) to find $Q_\xi ( \alpha )$. For example, the BTZ black hole with mass $M$ is a solution to the three-dimensional equations of motion for pure gravity (with negative cosmological constant) for any value of $M$. We can compute the change in the charge associated with the exact Killing vector $\xi = \partial_t$ in these backgrounds using the known expression for $\bk_\xi$ in Einstein gravity. The resulting integral gives
\begin{align}\label{change_in_charge_M}
    \frac{d Q_\xi}{d M} &= \oint_S \bk_\xi [ \delta_M g ;g ] \, ,\nonumber \\
    &= 1 \, .
\end{align}
Using the initial condition $Q_\xi ( M = 0 ) = 0$, we can then trivially integrate to find
\begin{align}
    Q_\xi = M \, ,
\end{align}
which recovers the fact that the mass of the black hole is the conserved charge associated with the global timelike Killing vector $\partial_t$.

\subsubsection*{\ul{\it Charge Integrability}}

Given a family of spacetimes labeled by some parameter $\alpha$, and possessing some exact Killing vector $\xi$, we have seen that the covariant phase space formalism allows us to compute the quantity $\frac{\partial Q_\xi}{\partial \alpha}$ which controls how the conserved charge associated with $\xi$ varies with $\alpha$. If $\alpha$ is the only parameter in our family of solutions, it is then trivial to integrate $\frac{\partial Q_\xi}{\partial \alpha}$ given an initial condition and recover the expression $Q_\xi ( \alpha )$ for the conserved charge.

Now consider a family of solutions which depend on several parameters $\alpha_1 , \ldots , \alpha_n$. The formalism permits us to compute a collection of partial derivatives $\frac{\partial Q_\xi}{\partial \alpha_i}$ for each $i$. We are not guaranteed that integrating any two of these quantities will yield equivalent charges. For instance, in the simple case of a solution for a black hole which depends on a mass parameter $M$ and a spin parameter $J$, we could obtain two separate derivatives
\begin{align}\begin{split}\label{integrability_kerr}
    \frac{\partial Q_\xi}{\partial M} &= \oint_S \bk_\xi [ \partial_M g ; g ] \, ,  \\
    \frac{\partial Q_\xi}{\partial J} &= \oint_S \bk_\xi [ \partial_J g ; g ]  \, .
\end{split}\end{align}
If computing $\partial_J \partial_M Q_\xi$ using the first line of (\ref{integrability_kerr}) yields the same result as computing $\partial_M \partial_J Q_\xi$ using the second line of (\ref{integrability_kerr}), then we can unambiguously define the charge $Q_\xi ( M, J )$ in a way which does not involve any choice of how to perform the integration. In this case, the charge $Q_\xi ( M , J )$ is said to be integrable. More generally, the integrability condition will hold so long as
\begin{align}
    \delta_1 \oint_S \bk_\xi [ \delta_2 \Phi ; \Phi ] = \delta_2 \oint_S \bk_\xi [ \delta_1 \Phi ; \Phi ] \, , 
\end{align}
for any pair of variations $\delta_1 \Phi$, $\delta_2 \Phi$.

Charge integrability holds in many cases of physical interest, such as the Kerr black hole in $D > 3$ dimensions or the spinning BTZ black hole in $D = 3$, both of which possess charges that are integrable in the space of $M$ and $J$ parameters. However, it was pointed out in \cite{Dias:2019wof} that solutions to the equations of motion for type IIB supergravity with multiple parameters $\alpha_i$ do not, in general, possess manifestly integrable charges. In such cases, if we are given only the information about the gravity solution, the definitions of conserved charges like mass and spin can be ambiguous; one obtains different answers depending on which quantity $\frac{\partial Q_\xi}{\partial \alpha_i}$ one integrates.

To resolve this ambiguity, one requires additional input which identifies a preferred notion of the surface charges. For instance, in the cases analyzed in \cite{Dias:2019wof}, the additional input comes from holography: if a type IIB supergravity solution is dual to a conformal field theory, one can use the scale invariance of the field theory to identify particular choices of the gravitational charges that are natural from the perspective of the boundary theory.

However, for more general solutions that are not dual to CFTs, there does not appear to be a general principle for constructing unambiguous charges in the case of non-integrable solutions. We will see shortly that the supergravity solutions constructed in section \ref{sec:generalized_solutions} possess exactly this ambiguity when $\Jt \neq 0$ . Because these bulk geometries are believed to be dual to field theories which are obtained through deforming a CFT by an irrelevant operator -- namely, something akin to a single-trace $\TT$ operator -- we cannot rely on the scale invariance of the dual theory to construct a preferred set of gravitational charges. 

\subsubsection*{\ul{\it Surfaces Charges in Type IIB-Like Theories}}

The formalism reviewed above applies to any generally covariant Lagrangian $\bL$. Next we will restrict attention to a class of Lagrangians which are relevant for type IIB supergravity theories in Einstein frame and with pure NS flux. Consider an action of the form
\begin{align}\label{general_lagrangian}
    S = \frac{1}{16 \pi G} \int d^D x \, \sqrt{-g} \left( R - \frac{1}{2} \partial_\mu \Phi \partial^\mu \Phi - \frac{1}{2} f ( \Phi ) | H |^2 \right) \, .
\end{align}
Here $\boldsymbol{H} = d \boldsymbol{B}$ is the field strength associated with a now general $p$-form potential $\boldsymbol{B} = B_{\mu_1 \ldots \mu_p} \, dx^{\mu_1} \ldots dx^{\mu_p}$, $|H|^2 = \boldsymbol{H} \wedge \ast \boldsymbol{H}$, and $f(\Phi)$ is an arbitrary function which controls the scalar coupling to the $p$-form kinetic term. When $D = 10$, $p = 2$, and $f ( \Phi ) = e^{- \Phi}$, this gives the Einstein-frame action for type IIB supergravity. After performing the conformal mapping from string-frame to Einstein-frame, the general solutions which we derived in section \ref{sec:generalized_solutions} satisfy the equations of motion associated with this action.

The surface charges associated with the Lagrangian (\ref{general_lagrangian}) were worked out in \cite{Compere:2007vx}; here we will recall the results which are relevant for our analysis.\footnote{These formulas have also been implemented in Mathematica by the author of \cite{Compere:2007vx} in a convenient package which may be found \href{https://ptm.ulb.be/gcompere/package.html}{here}.}

The contribution from the scalar is
\begin{align}\label{scalar_k_form}
    \bk^\Phi_\xi [ \delta g ,  \delta \Phi ; g ; \Phi ] = i_\xi \bTheta_\Phi \, , \qquad \bTheta_\Phi = \ast ( d \Phi \, \delta \Phi ) \, .
\end{align}
In components, (\ref{scalar_k_form}) is
\begin{align}
    k^{\mu \nu, \Phi}_{\xi} [ \delta g ,  \delta \Phi ; g ; \Phi ] = 2 \left( \delta \Phi \right) \cdot \xi^{[\nu} \partial^{\mu]} \Phi ,
\end{align}
where $T^{[ab]} = \frac{1}{2} \left( T^{ab} - T^{ba} \right)$ denotes the usual antisymmetrization. Recall that the non-bolface $k^{\mu \nu}_\xi$ is the antisymmetric $2$-tensor whose Hodge dual is the $(n-2)$-form $\bk_\xi$.

The contribution from the $p$-form $\boldsymbol{B}$ is more complicated and can be expressed as
\begin{align}\label{flux_k_form}
    \bk^{B}_{\xi} [ \delta g, \delta B ; g, B, \Phi ] = - \delta \bQ^{B}_{\xi} + i_\xi \bTheta_{B} - \bE_{\mathcal{L}}^{B} [ \mathcal{L}_\xi B , \delta B ] \, , 
\end{align}
where
\begin{align}
    \bQ^B_{\xi} &= f ( \Phi ) \left( i_\xi B \right) \wedge \ast H \, , \nonumber \\
    \bTheta^B &= f ( \Phi ) \left( \delta B \right) \wedge \ast H \, , \nonumber \\
    \bE_{\mathcal{L}}^{B} [ \mathcal{L}_\xi B , \delta B ] &= f ( \Phi ) \ast \left( \frac{1}{2 ( p - 1 ) !} \delta B_{\mu \alpha_1 \cdots \alpha_{p-1}} \tensor{\left( \mathcal{L}_\xi B \right)}{_\nu^{\alpha_1}^{\ldots}^{\alpha_{p-1}}} \, dx^\mu \wedge dx^\nu \right) \, .
\end{align}
It is convenient to separate the component expression for $k_\xi^{B, \mu \nu}$ into its contributions from the tensor field variation $\delta \boldsymbol{B}$ (and its field strength variation $\delta \boldsymbol{H} = d \left( \delta \boldsymbol B \right)$), the metric variation $\delta g_{\mu \nu}$, and the dilaton variation $\delta \Phi$:
\begin{align}
    k_\xi^{B, \mu \nu} &= 2 K_{\delta B}^{[\mu \nu]} + 2 K_{\delta g}^{[\mu \nu]} + 2 K_{\delta \Phi}^{[\mu \nu]} \, , \nonumber \\
    K_{\delta B}^{\mu \nu} &= - f ( \Phi ) \Bigg[ \frac{2}{p} \xi^\mu H^{\nu \alpha_1 \cdots \alpha_p } \left( \delta B \right)_{\alpha_1 \cdots \alpha_p} - \left( \mathcal{L}_\xi B \right)^{\mu \alpha_1 \cdots \alpha_{p-1}} \tensor{\left( \delta B \right)}{^\nu_{\alpha_1}_{\cdots}_{\alpha_{p-1}}}  \nonumber \\
    &\qquad \qquad \quad + H^{\mu \nu \alpha_1 \cdots \alpha_{p-1}} \xi^{\rho} \left( \delta B\right)_{\rho \alpha_1 \cdots \alpha_{p-1}} 
  + \left( \delta H \right)^{\mu \nu \alpha_1 \cdots \alpha_{p-1}} \xi^\rho B_{\rho \alpha_1 \cdots \alpha_{p-1}} \Bigg] \, , \nonumber \\ 
    K_{\delta g}^{\mu \nu} &= f ( \Phi ) \Bigg[ - \frac{1}{2} \tensor{\left( \delta g \right)}{^\alpha_\alpha} H^{\mu \nu \alpha_1 \cdots \alpha_{p-1}} \xi^{\rho} B_{\rho \alpha_1 \cdots \alpha_{p-1}} + ( p - 1 ) \left( \delta g \right)^{\mu \rho} \tensor{H}{_\rho^\nu^{\alpha_1}^{\cdots}^{\alpha_{p-1}}} \xi^\sigma B_{\sigma \alpha_1 \cdots \alpha_{p-1}} \nonumber \\
    &\qquad \qquad \quad + 2 \tensor{\left( \delta g \right)}{^\mu_\rho} H^{\rho \nu \alpha_1 \cdots \alpha_{p-1}} \xi^\sigma B_{\sigma \alpha_1 \cdots \alpha_{p-1}} \Bigg] \, , \nonumber \\
    K_{\delta \Phi}^{\mu \nu} &= - \frac{\partial f}{\partial \Phi} \left( \delta \Phi \right) H^{\mu \nu \alpha_1 \cdots \alpha_{p-1}} \xi^\rho B_{\rho \alpha_1 \cdots \alpha_{p-1}} \, .
\end{align}
Here we write $\delta g = g^{\mu \nu} \delta g_{\mu \nu}$ for the trace of the metric fluctuation.\footnote{This differs from the usage in expressions like $\bk_\xi^g [ \delta g; g ]$, where $\delta g$ refers to the full metric fluctuation rather than its trace, but where we have suppressed indices for ease of notation.} 

Finally, the contribution from the Einstein-Hilbert term $\sqrt{-g} R$ is
\begin{align}\label{metric_k_form}
    \bk_\xi^g [ \delta g; g ] = - \delta \bQ_\xi [ g ] - i_\xi \bTheta [ \delta g ; g ] \, , 
\end{align}
where
\begin{align}
    \Theta^\mu &= \frac{\sqrt{-g}}{16 \pi G} \left( \nabla_\nu \, \delta g^{\mu \nu} - \nabla^\mu \delta g \right) \, , \nonumber \\
    \bQ_\xi &= \frac{\sqrt{-g}}{8 \pi G} \nabla^\mu \xi^\nu \left( d^{D-2} x \right)_{\mu \nu} \, .
\end{align}
Here we follow the conventions of \cite{Compere:2018aar} for differential forms; for instance, $\left( d^{n - p} x \right)_{\mu_1 \ldots \mu_p}$ is defined by $\frac{1}{p! ( n - p )!} \varepsilon_{\mu_1 \ldots \mu_p \nu_{p+1} \ldots \nu_n} dx^{\nu_{p+1}} \wedge \ldots \wedge dx^{\nu_n}$ where $\varepsilon$ is the Levi-Civita symbol with entries $-1, 0$, or $1$ (without the factor of $\sqrt{-g}$). In components, (\ref{metric_k_form}) can be written as
\begin{align}
    k^{g, \mu \nu}_\xi = & \xi^{[\nu} \nabla^{\mu]} \delta g - \xi^{[\nu} \nabla_\alpha \delta g^{\mu] \alpha} + \xi_\alpha \nabla^{[\nu} \delta g^{\mu] \alpha} + \frac{1}{2} \delta g \nabla^{[ \nu} \xi^{\mu]} \cr & - \frac{1}{2} \delta g^{\alpha [\nu} \nabla_\alpha \xi^{\mu]} + \frac{1}{2} \delta g^{\alpha [\nu} \nabla^{\mu]} \xi_\alpha \, . 
\end{align}
The total contribution to the change in a charge $Q_\xi$ as a parameter $\alpha$ of the solution is varied, therefore, is given by
\begin{align}\label{overall_integral}
    \frac{\partial Q_\xi}{\partial \alpha} &= \oint_S \bk_\xi [ \partial_\alpha g ; g ] \, , \nonumber \\
    \bk_\xi &= \bk_\xi^{\Phi} + \bk_\xi^{B} + \bk_\xi^{g} \, .
\end{align}
We will take the surface $S$ to lie at a fixed time and fixed large value $R$ of the radial coordinate $r$ in our solutions, and then take the limit as $R \to \infty$. The integral (\ref{overall_integral}) then extracts the component $k_\xi^{t r}$ of the antisymmetric $2$-tensor which is Hodge dual to $\bk_\xi$. For the purpose of computing the integral (\ref{overall_integral}), the compact directions $S^3 \times T^4$ are merely spectators which are integrated over to yield an appropriate volume factor.

\subsection{Evaluating the Charge}\label{sec:non-rotating}

Our goal is to compute the Noether-Wald surface charge $Q_\xi$ associated with the exact Killing vector $\xi = \partial_t$ of the solutions with ALD asymptotics $(k_5=0)$ derived in section \ref{sec:generalized_solutions}. We will restrict to the case given in \C{final_string_frame_soln_bulk} of a non-rotating black hole with $\Jt = 0$. We will comment on the case with both mass and spin at the close of this section. That case involves interesting subtleties related to the non-integrability discussed in section \ref{sec:charge_review}. 

As is typical in the covariant phase space formalism, rather than computing the charge $Q_\xi$ directly, we first find $\frac{\partial Q_\xi}{\partial \alpha}$ where $\alpha$ is one of the parameters in these solutions. At this stage we have a choice, since our spacetimes depend on the parameters $(k_1, \Mt)$ with quantized fixed charges determined by $(m_1, m_5)$. If we vary both $(k_1, \Mt)$ then we find that the charge $Q_\xi$ is \emph{not} integrable in the space of these parameters; that is, we obtain different expressions for the mass if we, for instance, (1) compute $\frac{\partial Q_\xi}{\partial k_1}$ and integrate with respect to $k_1$, or (2) compute $\frac{\partial Q_\xi}{\partial \Mt}$ and integrate with respect to $\Mt$. In this case, we view the failure of integrability as reflecting the definition of an asymptotically linear dilaton spacetime. Namely, the parameter $k_1$ is a part of the definition of the asymptotic behavior and should not be varied.

For this reason, we will consider a one-parameter family of solutions where $\a_e$, which controls $\Mt$, is varied but all other parameters are held fixed. Therefore we will use the variations of the fields
\begin{align}\label{re_variations}
    \delta_{\alpha_e} g_{\mu \nu} = \frac{\partial g_{\mu \nu}}{\partial \alpha_e}  \, , \qquad \delta_{\alpha_e} B_{\mu \nu} = \frac{\partial B_{\mu \nu}}{\partial \alpha_e} \, , \qquad \delta_{\alpha_e} \Phi = \frac{\partial \Phi}{\partial \alpha_e}  \, .
\end{align}
Under this linearized variation, the combined contribution from $k_\xi^{g}$, $k_\xi^{B}$, and $k_\xi^{\Phi}$ to the change in the charge is
\begin{align}\label{charge_variation}
    \delta_{\alpha_e} Q_\xi = \oint_S \left( \bk_\xi^{g} + \bk_\xi^{B} + \bk_\xi^{\Phi} \right) = \frac{\rh_1 \alpha_e}{4 \sqrt{ \rh_1^2 + k_1 \alpha_e^2}}  \, .
\end{align}
Here $S$ is a fixed time slice at large $r$ in $\mathcal{M}_3$ and includes an integral over all the compact directions. Because we have used the variations (\ref{re_variations}) associated with changing $\a_e$, we have $\delta_{\alpha_e} Q_\xi = \frac{\partial Q_\xi}{\partial \a_e}$ so that
\begin{align}\label{mass_charge_derivative}
    \frac{\partial Q_\xi}{\partial \alpha_e} = \frac{\rh_1 \alpha_e}{4 \sqrt{ \rh_1^2 + k_1 \alpha_e^2}} \, .
\end{align}
This equation can be trivially integrated to find
\begin{align}
    Q_\xi = \frac{\rh_1}{4 k_1} \sqrt{ \rh_1^2 + k_1 \alpha_e^2} + C \, , 
\end{align}
where $C$ is an integration constant which may be a function of $k_1$ and $\alpha_1$ but not of $\alpha_e$. We fix the integration constant by requiring that the conserved charge $Q_\xi$ have a finite limit as $k_1 \to 0$. This is only possible if the constant $C$ is chosen so that the charge takes the form
\begin{align}
    Q_\xi = \frac{\rh_1^2}{4 k_1} \left( \sqrt{ 1 + k_1 \frac{\alpha_e^2}{\rh_1^2}} - 1 \right) \, .
\end{align}
To recover the undeformed black hole, we take $k_1 \to 0$. This is an easy way to fix the overall normalization. We will choose to normalize the charge, denoted $\widetilde{Q}_\xi $, so that $k_1\rightarrow 0$ gives $\Mt$: 
\begin{align}\label{sqrt_charge_soln}
    \widetilde{Q}_\xi = \frac{R^2}{4 r_5^2 k_1} \left( \sqrt{ 1 + k_1 \frac{8\Mt r_5^2}{R^2}} - 1 \right) \, .
\end{align}
This square root is {\it identical} to the expression that appears in the dilaton solution found in \C{final_string_frame_soln_bulk}. It is strong evidence for a holographic correspondence in the spirit proposed in~\cite{Giveon:2017nie}.

A straightforward extension of this analysis to the case with mass and spin encounters the issue of non-integrability in the space of $(\Mt, \Jt)$ solutions. Similar issues have been seen in past work~\cite{Dias:2019wof,Ruzziconi:2020wrb,Fiorucci:2021pha}. It would be interesting to resolve the non-integrability and see if one can reproduce the mass and charge for these asymptotically linear dilaton spacetimes expected from the form of the dilaton solution seen in \C{spinning_case}. 

\subsection{Winding Tachyon} \label{tachyon}

We have seen that the solution which interpolates from global $\AdS_3$ to ALD asymptotics, presented in section \ref{sec:generalized_solutions} with $\Mt = -\frac{1}{8}$, has a spacetime mass \C{sqrt_charge_soln} which becomes complex when $k_1$ is made too large. 

What might be happening in string theory when this ground state approaches this instability? Note that the string-frame metric sees an approximately constant circle for large $r$ of size $\frac{R}{\sqrt{k_1}}$ and a very weak string coupling. Global $\AdS$ in the interior corresponds to anti-periodic boundary conditions for spacetime fermions on the $x_5$ circle. Boundary conditions of this type tend to produce tachyons for strings wrapping the circle if the circle becomes too small~\cite{ROHM1984553}. Qualitatively, this is the kind of phenomenon we might hope would happen prior to the spacetime becoming unstable, although the endpoint of condensing such a bulk closed string tachyon is often non-geometric~\cite{Martinec:2009ks}. If the holographic theory exists for $k_1$ beyond the value where the ground state energy becomes complex, we will see that the more likely resolution lies in the low-energy gravity theory. 

To estimate where a tachyon might emerge, we note that a fundamental string on a circle of radius $\frac{R}{\sqrt{k_1}}$ with winding $w$ and a twisted boundary condition breaking supersymmetry satisfies a mass-shell condition,
\begin{align}\label{simplecircle}
    m^2 = \frac{w^2 R^2}{k_1 (\a')^2} + \frac{2}{\alpha'}\left( N_L + N_R \right)\, ,
\end{align}
for bosonic excitations with no momentum on the circle. In the $(NS-, NS-)$ sector with odd winding and no oscillators, the zero point energy is given by $N_L=N_R= - \frac{1}{2}$. The first state to become tachyonic has $w=1$ which happens for $k_1 > \frac{R^2}{2\alpha'}$. This is not quite right in our case because we have not taken the linear dilaton background into account. The linear dilaton slope is often denoted $Q$ in the literature on worldsheet descriptions of linear dilaton backgrounds, where in this case $Q=\frac{2}{\sqrt{m_5 \alpha'}}$; see, for example~\cite{Kutasov:2001uf}. For a string with no radial momentum, the mass-shell condition for a singly wound string is modified slightly from the simple circle of \C{simplecircle}, 
\begin{equation}
    m^2 = \frac{ R^2}{k_1 (\a')^2} + \frac{Q^2}{4}- \frac{2}{\alpha'}\, .
\end{equation}
Now the condition to avoid a tachyon becomes 
\begin{align}
    k_1 < \frac{R^2}{\a' \left( 2 - \frac{1}{m_5}\right)}\, .
\end{align}
This should be contrasted with the condition \C{gravity_condition} needed to avoid a complex dilaton solution, which we repeat here for convenience:
\begin{align}
    k_1 < \frac{R^2}{m_5 \a'} \, .
\end{align}
These conditions only agree for $m_5=1$. For large $m_5$ the gravity condition is violated first, long before one expects a closed string tachyon, suggesting a possible resolution in the low-energy theory without stringy ingredients. There is one final observation which is curious: if there were a `long' string with tension reduced by a factor of $m_5$ so the tension $T_{\rm long} = \frac{1}{2\pi \a' m_5}$ then a tachyon for this string would emerge before the gravity solution becomes complex as $k_1$ is increased.\footnote{Such long strings are expected from the low-energy physics of the D1-D5 system but are not expected to be visible in the weakly coupled perturbative string spectrum of the F1-NS5 system.} 

\section{The Two Parameter Family of Solutions}

Now we will explore the family of solutions that depend on an additional parameter $c_1$, which are derived in Appendix \ref{app:derivation_details}. 
These solutions are quite intriguing but exhibit some surprising features so our discussion is oriented around addressing the more basic issues with these solutions. 
At short distances and small black hole mass, the solutions again look like $\AdS_3$ black holes. At large distances, these solutions 
exhibit positive curvature in string-frame, which is typically hard to engineer in string theory in stationary backgrounds. 

With the details of the analysis again relegated to the Appendix, we first list the form of the string-frame metric for the case without spin: 
\begin{align}\label{positive_curvature_metric}
    & ds^2 = - \frac{1 - \frac{ \alpha_e^2}{r^2}}{k_1 + \frac{\a_1^2}{r^2}} \, dt^2 + \frac{1}{k_1 + \frac{\a_1^2}{r^2}} \, dx_5^2 + \frac{r_5^2 ( 1 + k_1 \alpha_1^2 c_1 ) ( 1 + k_1 c_1 ( \alpha_1^2 + k_1 \alpha_e^2 ) ) }{ ( r^2 - \alpha_e^2 ) ( 1 + k_1 c_1 ( \alpha_1^2 + k_1 r^2 ) )^2 }  \, dr^2\nonumber \\ & + r_5^2 \, d \Omega_3^2 + ds^2_{T^4} \, ,
\end{align}
with $\alpha_e^2 = 8 \Mt \frac{\rh_1^2 r_5^2}{R^2}$, and where the dilaton is given by
\begin{align}\label{c1_dilaton}
    e^{2\Phi} = c_2^2  \left( \frac{1+ k_1 c_1 \left( \alpha_1^2+k_1 r^2  \right)}{\alpha_1^2+k_1 r^2} \right) \, .
\end{align}
Notice that $e^{2\Phi}$ given in \C{c1_dilaton} always decreases as $r$ ranges from $0$ to $\infty$. Here the parameters $c_1$ and $c_2$ are related by the algebraic constraint
\begin{align} \label{c2algebraic}
    c_2^4 =  \frac{r_5^4}{\rh_1^4} \left\{  \frac{ \alpha_1^4 + k_1 \alpha_1^2 \alpha_e^2 } {( k_1 c_1 \alpha_1^2 + 1 ) ( k_1 c_1 ( \alpha_1^2 + k_1 \alpha_e^2 ) + 1 ) } \right\} \, .  
\end{align}
The curvatures for this metric are non-singular and the dilaton is well behaved as long as $k_1>0$ and $c_1>0$. Notice that the dilaton interpolates from a constant at small $r$ to a different constant for very large $r$.  

What is quite peculiar about the metric \C{positive_curvature_metric} is the behavior of $f_r$ which determines the $g_{rr}$ component of the metric. For large $r$, 
\begin{align}
f_r \sim \frac{1}{r^6} \, . 
\end{align}
This means $r=\infty$ is now a point at finite distance. How should we treat this boundary? There are several possible approaches which might give physically interesting backgrounds: 
\begin{itemize}
\item One can try to continue the metric beyond $r=\infty$. 
\item One can try to impose boundary conditions at $r=\infty$ or at some continuation of the metric. This is similar in spirit to the holographic proposal of \cite{McGough:2016lol}. 
\item One can try to include a brane source at or beyond $r=\infty$. If such a source has a sensible string theory interpretation then the metric might admit a natural continuation which takes into account the stress-energy and charge of the brane. 
\end{itemize}
We will explore one of these possibilities.

\subsubsection*{\ul{\it Continuing the metric}}

At large $r$, it is natural to change to the radial variable $\rhot = \frac{1}{r^2}$.\footnote{We are denoting this  radial variable by $\rhot$ to avoid confusion with the $\rho$ that appears in equation (\ref{defrho}) and surrounding discussion. } The range of $\rhot$ is $(\rhot_0, 0)$ prior to any continuation. Here $\rhot_0$ is a cutoff which is sufficiently small so that we can ignore $O \hspace{-1pt} \left( \rhot \right)$ corrections. We are interested in the physics for $\rhot \rightarrow 0$ where the three-dimensional spacetime metric takes the simple form, 
\begin{align}\begin{split}\label{c1_nonzero_solution}
    ds^2 &= - \frac{1}{k_1} \, dt^2 + \frac{1}{k_1} \, dx_5^2 + \frac{1}{k_{\rhot}} \, d \rhot^2 + O \hspace{-1pt} \left( \rhot \right) \, , \\
    k_{\rhot} &= \frac{4 k_1^4 c_1^2}{r_5^2 ( 1 + k_1 \alpha_1^2 c_1 ) \left( 1 + k_1 c_1 \alpha_1^2 + k_1^2 c_1 \alpha_e^2 \right) } \, ,
\end{split}\end{align}
where $k_{\rhot}$ is a positive constant. At this point, one could imagine imposing a boundary condition at $\rhot=0$ in the spirit of \cite{McGough:2016lol}. At this boundary, the string coupling has a finite value:
\begin{align}
    e^{2\Phi} = k_1 c_1 c_2^2 \, .
\end{align}
Or we might attempt to continue the metric by taking $\rhot$ negative. The string coupling takes the form, 
\begin{align}
    e^{2\Phi} = c_2^2  \left( \frac{\rhot + k_1 c_1 \left( \rhot \alpha_1^2+k_1 \right)}{\rhot \alpha_1^2+k_1} \right) \, .
\end{align}
The string coupling will either diverge or go to zero as we make $\rhot$ more negative. If $k_1 c_1 \a_1^2>0$ then the string coupling always goes to zero before we hit a strong coupling singularity. We want $k_1$ positive otherwise the $dx_5^2$ term of \C{positive_curvature_metric} will flip sign at finite $r$ introducing a closed time-like curve. If $c_1$ is not positive then the dilaton will not make sense for finite $r$ so it appears we must encounter a point where gravity is shut off before we meet any potential strong coupling singularity. 

This is quite suggestive. Holography for linear dilaton spacetimes is motivated, in part, by gravity shutting off at radial infinity because the string coupling goes to zero. Here we meet a finite value of $\rhot$ at which the string coupling vanishes given by,
\begin{align} \label{stringvanishes}
    \rhot = - \frac{k_1^2 c_1}{1+ k_1 c_1 \a_1^2}\, .
\end{align}

\subsubsection*{\ul{\it Finding a linear dilaton coordinate}}

After some experimentation and with some hindsight, it is useful to first answer the following question before  continuing the metric \C{c1_nonzero_solution}: is there a new radial coordinate that makes the dilaton of \C{c1_dilaton} look like the expression with $c_1=0$? The answer turns out to be yes in some cases,
\begin{align} \label{changeofr}
    \fr = \sqrt{ \frac{r^2 ( 1 - c_1 k_1 \alpha_1^2 ) - c_1 \alpha_1^4 }{1 + c_1 k_1 ( k_1 r^2 + \alpha_1^2 ) } } \, ,
\end{align}
with the dilaton taking the form, 
\begin{align} \label{fracrdilaton}
    e^{2 \Phi} = \frac{c_2^2}{k_1 \fr^2 + \alpha_1^2} \, .
\end{align}
The expression \C{changeofr} for $\fr$ makes sense for sufficiently large $r$ and sufficiently small $c_1 k_1$; specifically $r^2 \geq \frac{c_1 \a_1^4}{1-c_1 k_1 \a_1^2}$ and $c_1 k_1 \a_1^2 < 1$. At this point, we can ask what bounds we should impose on $(c_1, k_1)$ so that the solution for the dilaton is real. There appears to be no upper bound on $c_1>0$ and the bound on $k_1$ from considering \C{c1_dilaton} for the case of $\a_e^2<0$ is the same as our earlier discussion with $c_1=0$.   

Let us assume these conditions are satisfied. We can always choose $r$ to be sufficiently large but making $c_1 k_1$ sufficiently small is not necessary based on what we have seen so far. Note that the range of $\fr$ is $(0, \fr_{\infty})$ where
\begin{align}
    \fr_{\infty}  = \sqrt{ \frac{ ( 1 - c_1 k_1 \alpha_1^2 ) }{c_1 k_1^2  } }\, ,
\end{align}
is finite. Plotting $\fr$ as a function of $r$ shows that it is monotonic asymptoting to the value $\fr_{\infty}$. The natural continuation in this variable is to allow $\fr$ to continue to infinity past the finite value $\fr_{\infty}$. The  metric expressed in this radial variable takes the form, 
\begin{align}
    & ds^2 = \frac{\alpha_e^2 - c_1 ( \alpha_1^4 + k_1 \alpha_1^2 \alpha_e^2 ) - \fr^2 ( 1 + c_1 k_1 (\alpha_1^2 + k_1 \alpha_e^2 ) )}{\alpha_1^2 + k_1 \fr^2 } \, dt^2 + \left( c_1 \alpha_1^2 + \frac{\fr^2}{\alpha_1^2 + k_1 \fr^2} \right) \, dx_5^2 \nonumber \\
    & + \frac{r_5^2 \fr^2 ( 1 + c_1 k_1 \alpha_1^2 ) ( 1 + c_1 k_1 ( \alpha_1^2 + k_1 \alpha_e^2 ) ) }{ ( c_1 \alpha_1^4 + \fr^2 ( 1 + c_1 k_1 \alpha_1^2 ) ) ( c_1 \alpha_1^4 + ( c_1 k_1 \alpha_1^2 - 1 ) \alpha_e^2 + \fr^2 ( 1 + c_1 k_1 ( \alpha_1^2 + k_1 \alpha_e^2 ) ) ) } \, d\fr^2 \, ,
\end{align}
which simplifies as $\fr \rightarrow\infty$ to
\begin{align}\label{asymptoticc11}
    ds^2 &= \left( - \left\{ \frac{1}{k_1} + c_1 \left( \alpha_1^2 + k_1 \alpha_e^2 \right)\right\}+ \frac{1}{k_1^2 \fr^2} \left( \alpha_1^2 + k_1 \alpha_e^2 \right) \right) \, dt^2 + \left( \frac{1}{k_1} + c_1 \alpha_1^2 - \frac{\alpha_1^2}{k_1^2 \fr^2} \right) \, dx_5^2 , \nonumber \\
    &+ \frac{r_5^2}{\fr^2} \, d\fr^2 \ +  O \left( \frac{1}{\fr^4} \right) \, .
\end{align}
There are a couple of observations worth making concerning \C{asymptoticc11}. First the $d\fr^2 $ has precisely the form we expect for a linear dilaton theory and the dilaton \C{fracrdilaton} has the right radial dependence to define a linear dilaton spacetime. However the $\fr$-independent term in the $dt^2$ coefficient appearing in \C{asymptoticc11} now depends on $\a_e^2$. Similarly the $dx_5^2$ leading metric is $c_1$ modified so the radius of the asymptotic circle in string-frame has changed. 

\subsubsection*{\ul{\it Making the asymptotic radial metric canonical}}

We have just seen that finding a coordinate in which the dilaton has the same asymptotic form as the $c_1=0$ case is possible when $c_1 k_1 \a_1^2<1$. However, we have yet to see any reason to restrict $c_1$ to this particular case. There is another approach we can take. Let us see if we can find a coordinate in which the $dr^2$ coefficient of the metric \C{positive_curvature_metric} takes the asymptotic form 
\begin{align}
    \frac{r_5^2 d\ffr^2}{\ffr^2 - \hat{\a}_e^2} =  r_5^2 d\left({\rm arctanh}(\sqrt{1 - \frac{\hat{\a}_e^2}{\ffr^2}}) \right)^2 \sim \frac{r_5^2 d\ffr^2}{\ffr^2} \left(1+ O(\frac{ \hat{\a}_e^2}{\ffr^2}) \right) \, ,
\end{align}
 for a new coordinate $\ffr$ with a horizon at some potentially new location $\hat{\a}_e$. We can then see how the dilaton depends on this new coordinate.

 As a first case, let us take $\a_e=0$ and no horizon in the extension so  $\hat{\a}_e=0$ as well.\footnote{We will assume that the $\a_e$ of \C{positive_curvature_metric} is finite. One might also consider the case where the coordinate patch parametrized by $r$ is actually {\it inside} the black hole horizon with the horizon located in the extension past $r=\infty$.} In this case we can take, 
 \begin{align}
    \frac{1}{ \ffr} &=   \frac{\sqrt{ 1 + k_1 c_1 ( \alpha_1^2 + k_1 r^2 )  }}{r}\, , \\
     & =  k_1\sqrt{c_1} + O \left( \frac{1}{r^2} \right) \, ,
 \end{align}
 where the second line is expanded near $r=\infty$. We have picked the sign so that small $r$ corresponds to small $\ffr$. This relation can be inverted giving, 
 \begin{align}
     r^2 = \frac{(1 + k_1 c_1 \alpha_1^2)\,\ffr^2}{1 - c_1 k_1^2\, \ffr^2} \, .
 \end{align}
The point $r=\infty$ corresponds to $\ffr = \frac{1}{k_1 \sqrt{c_1}}$. The continuation corresponds to $\ffr$ extending past this point to infinity.
The dilaton can now be expressed in terms of $\ffr$,
\begin{align} \label{c1depdil}
    \Phi &=  \frac{1}{2} \log \left[ \frac{c_2^2 \left( 1 + k_1 c_1 \alpha_1^2 \right)}{k_1 \ffr^2 + \alpha_1^2 } \right] \, .
\end{align}
This is indeed an asymptotically linear dilaton spacetime with a slope that is superficially determined by $r_5$ and $c_1k_1$. Beautifully, however, the constant $c_2^2 \left( 1 + k_1 c_1 \alpha_1^2 \right)$ appearing in \C{c1depdil} becomes $c_1$-independent for this case of $\a_e=0$ using the relation \C{c2algebraic} and we recover the same dilaton as the  $c_1=0$ case.  We can now examine the full metric expressed in terms of $\ffr$, 
\begin{align}
    ds^2 = \frac{1 + c_1 k_1 \alpha_1^2}{k_1 + \frac{\alpha_1^2}{\ffr^2}} \left( - dt^2 + dx_5^2 \right) + \frac{r_5^2}{\ffr^2} \, d \ffr^2 \, .
\end{align}
Up to a rescaling of time and a redefinition of the asymptotic $x_5$ circle size, this solution is now identical to the $c_1=0$ case. 

Now we can turn to the case of $\a_e > 0$ with $r> \a_e$. The change of variables in this case is given by, 
\begin{align}
    \ffr^2 = \frac{\hat{\a}_e^2}{1- \frac{(r^2 - \a_e^2)(  1 + k_1 c_1  \alpha_1^2)}{r^2(  1 + k_1 c_1 ( \alpha_1^2 + k_1 \alpha_e^2  ))}} \, .
\end{align}
This is invertible giving, 
\begin{align}
    r^2 =  \frac{\a_e^2 \ffr^2}{\hat{\a}_e^2 + \frac{  \a_e^2 k_1^2 c_1( \hat{\a}_e^2 - \ffr^2)}{(  1 + k_1 c_1  \alpha_1^2)}} \, .
\end{align}
Substituting into the dilaton expression \C{c1_dilaton} gives, 
\begin{align} \label{messierdil}
    e^{2\phi} = c_2^2  \left( \frac{ \hat{\a}_e^2( 1+ k_1 c_1 \alpha_1^2 )( 1 + k_1 c_1 ( \alpha_1^2 + k_1 \alpha_e^2 ))}{k_1 \a_e^2 \, \ffr^2 + (1+k_1c_1\a_1^2 + k_1^2 c_1 \a_e^2 ) \a_1^2 \hat{\a}_e^2 } \right) \, .
\end{align}
If we impose the condition that the denominator of \C{messierdil} take the form seen in \C{c1_dilaton} then we relate $\a_e$ and $\hat{\a}$ as follows:
\begin{align}
    \alpha_e = \hat{\alpha}_e \sqrt{ \frac{1 + k_1 c_1 \alpha_1^2}{1 - k_1^2 c_1 \hat{\alpha}_e^2} } \, .
\end{align}
On substituting back into the dilaton we wonderfully find all factors conspire to give,
\begin{align}
    e^{2 \Phi} =  \frac{r_5^2}{\rh_1^2} \left\{  \sqrt{ \alpha_1^4 + k_1 \alpha_1^2 \hat{\alpha}_e^2  } \right\} \cdot \frac{1}{\alpha_1^2 + k_1 \ffr^2} \, , 
\end{align}
which is the form expected for a solution with $c_1=0$. At least for these cases, analytic continuation together with a suitable map of parameters recovers the solutions discussed earlier with $c_1=0$.

\section*{Acknowledgements}

It is our pleasure to thank D.~Kutasov, E.~Martinec and R.~Wald for helpful discussions. C.\,C., C.\,F. and S.\,S. are supported in part by NSF Grants No. PHY1720480 and PHY2014195. C.~F. is also supported by U.S. Department of Energy grant DE-SC0009999 and by funds from the University of California.

\appendix\newpage

\section{Derivation of General Solutions}\label{app:derivation_details}

In this Appendix we derive a general class of type IIB supergravity solutions which satisfy the assumptions laid out in section \ref{subsec:description_general_solution}. We will work in string frame with the action (\ref{IIB_action}), for which the associated equations of motion for the metric, dilaton, and flux are
\begin{align}
    R_{\mu \nu} + 2 \nabla_\mu \nabla_\nu \Phi - \frac{1}{4} | H |^2_{\mu \nu} = 0 \ , \label{einstein_eqn} \\
    R + 4 \nabla^2 \Phi - 4 | \nabla \Phi |^2 - \frac{1}{12} | H |^2 = 0 \, , \label{dilaton_eom} \\
    \nabla^\mu \left( e^{-2 \Phi} H_{\mu \nu \rho} \right) = 0 \, , \label{flux_eom}
\end{align}
respectively. A combination of the first two equations gives a nice equation for the dilaton. See, for example~\cite{Maxfield:2014wea}: 
\begin{align} \label{dillaplacian}
    \nabla^2 e^{-2\Phi} = e^{-2\Phi} |H|^2. 
\end{align}

We will begin with the following general ansatz consistent with the symmetry and flux assumptions stated in section~\ref{subsec:description_general_solution}:
\begin{align}\label{black_hole_ansatz}
    ds^2 &= - \frac{f_e ( r )}{f_1 ( r ) } \, dt^2 + \frac{1}{f_x ( r ) } \, \left\{ d x_5 + f_j ( r ) \, dt \right\}^2 + f_r ( r )  \, dr^2 + r^2 f_5 ( r )  \, d \Omega_3^2 + f_{T^4} ( r ) ds^2_{T^4} \, , \nonumber \\
    &\hspace{70pt} \Phi = \Phi ( r ) \, , \qquad H_3 =  \frac{2 m_5}{\ell_s} \epsilon_{S^3}  + f_{\mathcal{M}_3} ( r ) e^{2 \Phi} \epsilon_3^{\mathcal{M}_3} \, .
\end{align}
Here $\e_{S^3}$ is the volume form for $S^3$ with $\int \e_{S^3} = 2\pi^2 \ell_s^3$. Flux quantization through $S^3$ is automatically satisfied. Flux quantization through $\M_3$ requires that 
\begin{align}\label{fM3}
    f_{\mathcal{M}_3} ( r ) = \frac{32 m_1 \pi^4 \alpha^{\prime 3} g_s^2}{V_4 r^3 f_5 ( r )^{3/2}} .
\end{align}
That fixes one unknown function. This form for $H_3$ also solves \C{flux_eom}. One way to make $R_{\m\n}$ vanish in the $T^4$ directions is by setting $f_{T^4}(r)$ to a constant. It is not clear that any other solution exists so will set $f_{T^4}(r)=1$. This is in accord with the torus playing no significant role in the physics. 

The choice of $f_r$ is the freedom to parametrize the radial direction in a convenient way. 
At this point, we need to understand what asymptotic conditions to impose on our spacetime at small $r$ and large $r$.

\subsection{The partially decoupled case $k_5=0$}

\subsubsection*{\ul{\it Interpolating ansatz}}

It will simplify our analysis to take the following ansatz for the ten-dimensional metric,
\begin{align}\begin{split} \label{metricansatzk5=0}
    & f_1(r) = k_1 + \frac{\alpha_1^2}{r^2}, \qquad  f_e(r) = 1 - \frac{ \alpha_e^2}{r^2}, \qquad f_5(r)  = \frac{\alpha_5^2}{r^2},  \\
    & f_x(r) = k_x + {\alpha_x^2 \over r^2}  \, , \qquad  f_j(r)=0 \, , 
\end{split}\end{align}
 with $f_j$ set to $0$ because $J=0$. We could have considered the more general $f_5(r)  = k_5+ \frac{\alpha_5^2}{r^2}$ but for this partially decoupled case, we set $k_5=0$. We also demand that $k_1>0$ and $k_x>0$ for a sensible metric at large $r$. We have chosen to absorb the constant $k_e$ that might have appeared in $f_e$ into a redefinition of the other parameters. Because this is the string-frame metric rather than the Einstein-frame metric, we will leave $k_1$ in the ansatz until later when we determine how to normalize the time direction at large $r$. 
 
 \subsubsection*{\ul{\it Constraints from equations of motion}}

With this assumption, plugging our ansatz into the Einstein and dilaton equations gives constraints, 
\begin{align}\label{solveax}
    \alpha_x^2 = \frac{k_x }{k_1} \alpha_1^2 \, , \qquad \alpha_5 = r_5 \, , 
\end{align}
and the following solution for  $e^{2\Phi}$,
\begin{align} \label{dilatonsolution}
    e^{2\Phi} = c_2^2  \left( \frac{1+ k_1 c_1 \left( \alpha_1^2+k_1 r^2  \right)}{\alpha_1^2+k_1 r^2} \right) \, , 
\end{align}
where $c_1$ and $c_2$ are unknown constants. Imposing the dilaton equation of motion relates $c_1$ and $c_2$ as follows,
 \begin{align}\label{relatec1c2}
    \frac{r_5^4}{\rh_1^4} - \frac{( k_1 c_1 \alpha_1^2 + 1 ) ( k_1 c_1 ( \alpha_1^2 + k_1 \alpha_e^2 ) + 1 ) }{ \alpha_1^4 + k_1 \alpha_1^2 \alpha_e^2 } c_2^4  = 0 \, .
\end{align}
Note that $c_2$ cannot be set to zero. At this juncture, we have not imposed either small $r$ or large $r$ asymptotics on the dilaton. 
The constants $(c_1, c_2)$ determine the dilaton and $f_r$. Everything else is fixed by both matching onto BTZ at small $r$ and imposing large $r$ asymptotics, which we have yet to examine. 

If we now demand that $g_s \rightarrow 0$ as $r\rightarrow \infty$ then we need to set $c_1=0$ in~\C{dilatonsolution} to get a dilaton:
\begin{align}\label{dilc1=0}
    e^{2\Phi} =  \left( \frac{c_2^2 }{\alpha_1^2+k_1 r^2} \right) \, .  
\end{align}
We also require $\frac{c_2^2}{k_1}\geq 0$ since it is the coefficient of the large $r$ string coupling of \C{dilatonsolution} so $c_2^2\geq 0$. For a sensible value of $e^{2\Phi}$ in a neighborhood of $r=0$ we must also then require $\alpha_1^2>0$. This also fixes $c_2$ in terms of the other parameters of the ansatz using \C{relatec1c2}:
\begin{align} \label{constraintonc2}
    c_2^4 = \frac{r_5^4}{\rh_1^4} ( \alpha_1^4 + k_1 \alpha_1^2 \alpha_e^2 )  \, .
\end{align}
With the dilaton determined, we can now solve for $f_r$: 
\begin{align} \label{frc1=0}
    f_r ( r ) = \frac{ r_5^2}{ r^2- \alpha_e^2}\, .
\end{align}

\subsubsection*{\ul{\it Small $r$ asymptotics}}

At very small $r$ we demand that the solution in Einstein-frame look like a three-dimensional BTZ black hole. This expectation is at least reasonable for low mass black holes whose structure should not care very much about the large $r$ asymptotic behavior of the metric.
For simplicity let us first consider a black hole with mass but no spin. In the notation of the full decoupling limit presented in equation (\ref{full_decoupling}), the metric of a three-dimensional BTZ black hole of mass $\Mt$ and spin $J=0$ is
\begin{align}\label{BTZM=0}
    \widehat{ds}^2 = \ell^2 \left( - \left( - 8 \Mt + \rt^2 \right) d \tilde{t}^2 + \rt^2 d \tilde{\varphi}^2 + \frac{d \rt^2}{- 8 \Mt + \rt^2 } \right) \, ,
\end{align}    
where the mass parameter $\Mt$ has been made dimensionless using the effective three-dimensional Newton constant.

We will initially be noncommittal about the relation between these dimensionless variables and the dimensionful variables of our ansatz \C{metricansatzk5=0}. Because $\alpha_5=r_5$ by the equations of motion, the reduction on $S^3$ and $T^4$ to a three-dimensional theory is unchanged from the discussion around (\ref{full_decoupling}). There are no mass-dependent factors from integrating over those volumes in string-frame.

The three-dimensional Einstein-frame metric is given by, 
\begin{align}\label{einsteinframe}
    ds^2_{E} = e^{-4\Phi} ds^2_{\rm string} \, .
\end{align}
 The most commonly studied backgrounds involve a constant dilaton so moving between string-frame and Einstein-frame is straightforward. This is no longer immediately the case in backgrounds where the dilaton varies so we will want to keep careful track of dilaton factors. For this $c_1=0$ case we can expand $e^{-4\Phi}$ at small $r$ to find, 
\begin{align}
    e^{-4\Phi} = \frac{\alpha_1^4}{c_2^4} \left( 1 + \frac{2 k_1 r^2 }{\alpha_1^2}  \right) + O(r^4) \, , 
\end{align}
where $c_2^4$ is given in~\C{constraintonc2}.

 How shall we fix the small $r$ asymptotics? Let us first note that there is no singularity in $e^{2\Phi}$ which must be positive for all $r$ so the conformal factor in \C{einsteinframe} does not introduce any potential singularities. On the other hand, $f_r$ appearing in \C{frc1=0} does have a singularity at $r=\alpha_e$. Comparing with \C{BTZM=0} tells us that 
 \begin{align}
     \alpha_e^2 \sim \Mt \, .
 \end{align} 
We still need to relate the coordinates of our ansatz to the dimensionless coordinates of the BTZ solution. There is a natural place to begin such an identification with, 
\begin{align}
    e^{-4\Phi} {1\over f_x(r)} dx_5^2 \, \rightarrow \, \ell^2 \rt^2 d \tilde{\varphi}^2 \, ,
\end{align}
for small $r$. There is no freedom to redefine the angular coordinate which allows us to unambiguously identify:
\begin{align} \label{relatingr}
     e^{-4\Phi} r^2 = \ell^2 \frac{\alpha_x^2}{R^2 }  \, \rt^2 = \ell^2 \frac{k_x}{k_1} \frac{ \alpha_1^2}{ R^2} \,\rt^2    \quad \Rightarrow \quad r^2 = \ell^2 \, \frac{k_x}{k_1} \frac{ c_2^4}{ \alpha_1^2 R^2} \,\rt^2 +O(\rt^4) \, .
\end{align}
Once we have this relation between the dimensionless $\rt$ and $r$, we can determine a small $\rt$ regime where the  dilaton \C{dilc1=0} is approximately constant. This requires that $\frac{d\Phi}{d\rt}$ be small:
\begin{align}
    {k_x} \frac{\ell^2 c_2^4 }{\a_1^4 R^2} \rt  = \sqrt{k_1 k_x} \frac{\ell c_2^2}{\a_1^3 R} r \ll 1. 
\end{align}
In this regime we demand that the solution looks like a BTZ black hole in approximately $\AdS$ space.

Now we can examine the metric coefficient of $dr^2$ in \C{black_hole_ansatz}, 
\begin{align}
    e^{-4\Phi} f_r(r) dr^2  & = \left( \frac{\alpha_1^4}{c_2^4} \left( 1 + \frac{2 k_1 r^2 }{\alpha_1^2}  \right) + \ldots\right) \left( \frac{ r_5^2}{ r^2- \alpha_e^2}\right) dr^2 \, .
\end{align}
Comparison with  \C{BTZM=0} teaches us that 
\begin{align}\label{l_alpha_e_rel}
    \ell^2 = \frac{\alpha_1^4}{c_2^4} r_5^2, \qquad \alpha_e^2 = 8 \Mt \frac{k_x}{k_1}\frac{\alpha_1^2 r_5^2}{R^2}\, .
\end{align}
 As expected, $\alpha_e^2$ is related to the mass of the black hole. Interestingly the definition of $\ell$ looks mass-dependent when $k_1 \neq 0$ unlike the case of constant dilaton. 
We can also simplify \C{relatingr}, 
\begin{align}
    \frac{r^2}{\rt^2} = \frac{k_x}{k_1} \frac{\a_1^2 r_5^2}{R^2}, 
\end{align}
and the regime where the solution will look approximately like $\AdS$:
\begin{align}
\sqrt{k_1k_x} \frac{r_5 r}{\a_1 R} \ll 1. 
\end{align}

The remaining task is to relate $t$ and $\tilde{t}$, which follows by requiring the singularity at $r^2 = \alpha_e^2$ be a coordinate singularity. Let us recall that $e^{-4\Phi}\frac{1}{f_1(r)}$ is positive and introduces neither singularities nor changes to the signature of the time coordinate. On the other hand, $f_r(r)$ does change sign at $r=\a_e$ and this change must be matched by the behavior of the $dt^2$ metric coefficient. This constrains the $dt^2$ metric coefficient to match BTZ, 
\begin{align}\label{timecomponentofmetric}
   \frac{\alpha_1^2}{c_2^4} \left( {r^2 - \a_e^2} \right ) \left\{ 1+O(r^4) \right\} dt^2 = \ell^2 \left( \rt^2 - 8 \Mt \right) d\tilde{t}^2\, . 
\end{align}
From the $O(r^2)$ term we conclude:
\begin{align}
    d\tilde{t}^2 =  \frac{r^2}{\rt^2} \frac{\a_1^2}{c_2^4 \ell^2} \, dt^2 \quad \implies \quad  \tilde{t} = \frac{t}{R} \sqrt{\frac{k_x}{k_1}}\, .
\end{align}
The other $r$-independent condition from \C{timecomponentofmetric} is then automatically satisfied. 

\subsubsection*{\ul{\it Matching large and small $r$ asymptotics }}

To go further we need to examine the large $r$ asymptotics. In essence, the issue is what large $r$ criteria define metrics which are asymptotically linear dilaton? To understand the large $r$ asymptotics, we need an appropriate large $r$ coordinate. From the $e^{-4\Phi}f_r(r) dr^2$ term with the explicit form for $f_r$ given in \C{frc1=0}, we see that the natural radial coordinate for large $r$ takes the form,
\begin{align} \label{defrho}
    \rho = \frac{r_5 k_1}{2 c_2^2} r^2 = \frac{r^2}{\a_\rho} \, , \qquad \a_\rho = \frac{2 c_2^2}{r_5 k_1}\, . 
\end{align}
In terms of $\rho$, the three-dimensional asymptotic metric takes the form:
\begin{align}\label{asymptotic_metric}
    ds^2 &= - \frac{4\rho^2}{r_5^2}\left( \frac{1}{k_1} + O \left( \frac{1}{\rho} \right) \right)\, dt^2 + \frac{4\rho^2}{r_5^2}\left( \frac{1}{k_x} + O \left( \frac{1}{\rho} \right) \right) \, dx_5^2 + d\rho^2  \, .
\end{align}
What is important is that the leading term at large $\rho$ is completely independent of the mass $\Mt$. This defines the vacuum to which each solution should asymptote with mass-dependent subleading corrections.  

Our remaining task is to identify $\a_1^2$. We have one example of an asymptotically linear dilaton metric in \C{partialasymptotics} which corresponds to $\Mt=0$ BTZ. This case turns out to be somewhat special when compared with black holes with $\Mt \neq 0$. For example, any non-zero value of $k_1$ can be set to $1$ by a redefinition of the radial coordinate.\footnote{There is a similar special feature for $E=0$ states under a $\TT$ deformation; namely the energy does not flow. This is readily visible from \C{energyformula}. } This is not the case when $\Mt \neq 0$. Regardless, we can still try to use this case to normalize $\a_1^2$ in our metric ansatz \C{metricansatzk5=0}. Let us examine the large $\rho$  behavior of our dilaton: 
\begin{align}\label{asymptotic_dilaton}
    \Phi &= - \frac{1}{2}\log\rho + \frac{1}{2}\log \frac{r_5}{2} -  \frac{1}{2} \log\left(1 + \frac{ \a_1^2 r_5}{2 c_2^2} \frac{1}{\rho} \right) \, \nonumber\\
& =  - \frac{1}{2}\log\rho + \frac{1}{2}\log \frac{r_5}{2} - \frac{ \a_1^2 r_5}{4 c_2^2} \frac{1}{\rho} + O \left( \frac{1}{\rho^2} \right) \, .
\end{align}
All the potential mass-dependence, which is hidden in $c_2^2$, is $\rho$-suppressed. If we set $\a_e=0$ and compare these expressions for the dilaton and metric with the expressions found in \C{partialasymptotics}, we can identify $\a_1^2$:
\begin{align}
    \a_1^2 =  \rh_1^2 \, .
\end{align}
At this point we could redefine time to absorb the $\frac{1}{k_1}$ factor with the replacement $\a_1^2 \rightarrow \frac{\a_1^2}{k_1}$. However the $\AdS$ limit where $k_1 \rightarrow 0$ is more transparent with the explicit $k_1$.  On the other hand, redefining the periodicity of $x_5$ so that $k_x \rightarrow k_1$ does look natural. With this redefinition, $\a_x \rightarrow \a_1$ using \C{solveax}. The final string-frame metric and dilaton take the form, 
\begin{align}\label{final_string_frame_soln}
   &  ds^2 = - \frac{1 - \frac{ \alpha_e^2}{r^2}}{k_1 + \frac{\rh_1^2}{r^2}} \, dt^2 + \frac{1}{k_1 + \frac{\rh_1^2}{r^2}} \, dx_5^2 + \frac{ r_5^2}{ r^2- \alpha_e^2}  \, dr^2 + r_5^2 \, d \Omega_3^2 + ds^2_{T^4} \, , \nonumber \\
&  e^{2\Phi} = \frac{r_5^2 \sqrt{1 + 8 k_1\frac{ \Mt  r_5^2}{R^2} }}{\rh_1^2 + k_1 r^2} \, , \qquad \alpha_e^2 = 8 \Mt \frac{\rh_1^2 r_5^2}{R^2}\, .
\end{align}
Note that the $x_5$ circle  has periodicity $\frac{2\pi R}{\sqrt{k_1}}$ at $r=\infty$ in string-frame. 

\subsubsection*{\ul{\it The case of $c_1 \neq 0$}}

In solving for the dilaton, we set $c_1=0$ in \C{dilatonsolution}. In this subsection we will examine whether there are any sensible solutions with $c_1>0$. The case of $c_1<0$ is ruled out because there would be a singularity in the dilaton at some finite $r$. If a solution exists with $c_1>0$, it would be quite curious because it interpolates from one approximately constant value of the dilaton near $r=0$ to a different approximately constant value near $r=\infty$. 

The first step is to solve for $c_2^4$ using \C{relatec1c2}. This is just algebraic and gives:
\begin{align} \label{c2intermsofc1}
    c_2^4 =  \frac{r_5^4}{\rh_1^4} \left\{  \frac{ \alpha_1^4 + k_1 \alpha_1^2 \alpha_e^2 } {( k_1 c_1 \alpha_1^2 + 1 ) ( k_1 c_1 ( \alpha_1^2 + k_1 \alpha_e^2 ) + 1 ) } \right\} \, .  
\end{align}
The  $r\rightarrow \infty$ UV value of the dilaton is determined by  $e^{2\Phi} = k_1 c_1 c_2^2$ while the $r\rightarrow 0$ IR value,
\begin{align}
e^{2\Phi} =  \left( \frac{1+ k_1 c_1  \alpha_1^2}{\alpha_1^2} \right) c_2^2 >  k_1 c_1 c_2^2\, , 
\end{align}
is strictly larger. If the asymptotic solutions are both approximately $\AdS$ solutions with the same string-frame length scale $\ell$ then this relation would be in accord with the two-dimensional $c$-theorem. This suggests a potential holographic interpretation of this class of solutions as describing a flow from a UV two-dimensional CFT to an IR CFT. What is particularly interesting about this possibility is that by taking $c_1$ very small, this background would have a long intermediate region that looks like a linear dilaton but which eventually becomes $\AdS_3$ again.  

When $c_1 > 0$, $f_r$ is constrained to be
\begin{align}
    f_r ( r ) = \frac{r_5^2 ( 1 + k_1 \alpha_1^2 c_1 ) ( 1 + k_1 c_1 ( \alpha_1^2 + k_1 \alpha_e^2 ) ) }{ ( r - \alpha_e ) ( r + \alpha_e ) ( 1 + k_1 c_1 ( \alpha_1^2 + k_1 r^2 ) )^2 } \, ,
\end{align}
which diverges at $r = \alpha_e$ just like the $c_1 = 0$ solution. Again it appears that this is only a coordinate singularity and not a genuine curvature singularity. The full Ricci scalar associated with this ten-dimensional solution is given by
\begin{align}
    &R = \frac{2 k_1}{r_5^2} \times \nonumber \\
    &\frac{\left( 3 k_1 r^4 + 10 r^2 \alpha_1^2 + 2 k_1 r^2 \alpha_e^2 - 5 \alpha_1^2 \alpha_e^2 + 5 k_1 c_1 ( k_1 r^2 + \alpha_1^2 ) ( r^2 ( 2 \alpha_1^2 + k_1 \alpha_e^2 ) - \alpha_1^2 \alpha_e^2 ) \right) }{( k_1 r^2 + \alpha_1^2 )^2 ( 1 + k_1 c_1 \alpha_1^2 ) \left( 1 + k_1 c_1 ( \alpha_1^2 + k_1 \alpha_e^2 ) \right) }   \, .
\end{align}
If $k_1 > 0$ and $\alpha_1^2 > 0$, the Ricci scalar is finite for all values of $r$. It appears to interpolate from a constant negative value at small $r$ to a constant positive value at large $r$. One finds
\begin{align}
    \lim_{r \to 0} R = - \frac{10 k_1 \alpha_e^2}{r_5^2 \alpha_1^2 ( 1 + k_1 \alpha_1^2 c_1 + k_1^2 \alpha_e^2 c_1 ) } \, , 
\end{align}
while
\begin{align}
    \lim_{r \to \infty} R = \frac{2 \left( 3 + 10 k_1 \alpha_1^2 c_1 + 5 c_1 k_1^2 \alpha_e^2 \right) }{r_5^2 ( 1 + k_1 c_1 \alpha_1^2 ) ( 1 + k_1 \alpha_1^2 c_1 + k_1^2 \alpha_e^2 c_1 ) } \, .
\end{align}
We can also compute the Ricci scalar $R^{(3)}$ associated with the three-dimensional spacetime $\mathcal{M}_3$ parameterized by $t, r, x_5$. This is given by
\begin{align}\hspace{-15pt}
    R^{(3)} =\, &   2 \left( 1 + k_1 c_1 ( k_1 r^2 + \alpha_1^2 ) \right) \times \cr &  \frac{ \left( 4 k_1 r^2 \alpha_1^2 - 3 \alpha_1^4 + 2 k_1^2 r^2 \alpha_e^2 - 5 k_1 \alpha_1^2 \alpha_e^2 - 3 k_1 c_1 \alpha_1^2 ( k_1 r^2 + \alpha_1^2 ) ( \alpha_1^2 + k_1 \alpha_e^2 ) \right) }{r_5^2 ( k_1 r^2 + \alpha_1^2 )^2 ( 1 + k_1 c_1 \alpha_1^2 ) ( 1 + k_1 c_1 ( \alpha_1^2 + k_1 \alpha_e^2 ) ) } \, .
\end{align}
Again we can look at the small $r$ and large $r$ limits,
\begin{align}
    \lim_{r \to 0} R^{(3)} &= - \frac{2 \left( 3 \alpha_1^2 + 5 k_1 \alpha_e^2 + 3 k_1 c_1 \alpha_1^2 ( \alpha_1^2 + k_1 \alpha_e^2 ) \right) }{r_5^2 \alpha_1^2 ( 1 + k_1 c_1 ( \alpha_1^2 + k_1 \alpha_e^2 ) )} \, , \nonumber \\
    \lim_{r \to \infty} R^{(3)} &= \frac{2 c_1 \left( 4 k_1 \alpha_1^2 + 2 k_1^2 \alpha_e^2 - 3 k_1^2 c_1 \alpha_1^2 ( \alpha_1^2 + k_1 \alpha_e^2 )  \right) }{r_5^2 ( 1 + k_1 c_1 \alpha_1^2 ) ( 1 + k_1 c_1 ( \alpha_1^2 + k_1 \alpha_e^2 ) ) } \, .
\end{align}
We see that the small $r$ limit of $R^{(3)}$ is negative-definite, corresponding to an $\AdS_3$ in the deep interior, but that the large $r$ value of $R^{(3)}$ has competing contributions with opposite signs. In order to have negative curvature at large $r$, we require
\begin{align}
    c_1 > \frac{2}{3 k_1 } \left( \frac{1}{\alpha_1^2} + \frac{1}{\alpha_1^2 + k_1 \alpha_e^2 }  \right) \, .
\end{align}
Notice that this condition is easy to violate by taking $k_1$ small. The surprising result is that we seem to be able to construct a three-dimensional spacetime $\M_3$ with {\it positive} scalar curvature at least in string-frame! 

Our remaining task with these solutions is to determine the parameter map following the analysis of the $c_1=0$ case. We will highlight the differences. For simplicity, we assume $k_x=k_1$. The relation between $r^2$ and $\rt^2$ given in \C{relatingr} now becomes,
\begin{align} \label{relatingr_c1}
      r^2 = \ell^2 \, \frac{ c_2^4}{ \alpha_1^2 R^2} \,\frac{\rt^2}{(1+k_1 c_1 \a_1^2)^2} +O(\rt^4) \, .
\end{align}
Also from,  
\begin{align}
    e^{-4\Phi} f_r(r) dr^2  & = e^{-4\Phi} \,  \frac{ ( 1 + k_1 \alpha_1^2 c_1 ) ( 1 + k_1 c_1 ( \alpha_1^2 + k_1 \alpha_e^2 ) ) }{ \left( 1 + k_1 c_1 ( \alpha_1^2 + k_1 r^2 ) \right)^2 }  \, \frac{ r_5^2 \, dr^2}{r^2 - \alpha_e^2 } \, ,
\end{align}
we can read off
\begin{align}
    \ell^2 = \frac{\a_1^4 r_5^2}{c_2^4} \,  \frac{ ( 1 + k_1 c_1 ( \alpha_1^2 + k_1 \alpha_e^2 ) ) }{ \left( 1 + k_1 c_1  \alpha_1^2 \right)^3 } \, ,
\end{align}
which has additional mass dependence beyond the $c_1=0$ case. The last step is to relate $\a_e$ to $\Mt$ which now gives the following relations:
\begin{align}
    r^2 &= \frac{\a_1^2 r_5^2}{R^2} \frac{( 1 + k_1 c_1 ( \alpha_1^2 + k_1 \alpha_e^2 ))}{\left( 1 + k_1 c_1  \alpha_1^2 \right)^5} \, \rt^2\, , \\
    \a_e^2 &= \frac{8 \Mt r_5^2 \a_1^2 \left( 1 + c_1 k_1 \alpha_1^2 \right) }{R^2 \left( 1 + c_1 k_1 \alpha_1^2 \right)^5 - 8 c_1 k_1^2 \Mt r_5^2 \alpha_1^2} \, .
%
\end{align}

\subsubsection*{\ul{\it Adding spin $\Jt$}}

Now let us return to the case $c_1=0$ and extend our solutions \C{final_string_frame_soln} to include spin. This will also help us identify the relation between the deformation parameter $k_1$ and $\l$ of \C{energyformula}. The BTZ solution with spin takes the form, 
\begin{align}\label{BTZJ}
    \widehat{ds}^2 =& \ell^2 \left( - \left( - 8 \Mt + \rt^2 + \frac{16 \Jt^2}{ \rt^2} \right) d \tilde{t}^2 + \rt^2 \left(d \tilde{\varphi} - \frac{4 \Jt}{ \rt^2} d{\tilde t}\right)^2  \right. \cr & \left. + \frac{d \rt^2}{- 8 \Mt + \rt^2 + \frac{16 \Jt^2}{ \rt^2}} \right) \, .
\end{align} 
In parallel with the mass $\Mt = MG_3$, we have absorbed factors of the gravitational coupling into the angular momentum via $\Jt = \frac{J G_3}{\ell}$ with $G_3$ defined in \C{kappa3_defn}. This redefinition is quite important because we have already seen that $\ell$ is mass-dependent in (\ref{l_alpha_e_rel}). Soon we will see that it also becomes spin-dependent when $k_1 \neq 0$.  We will use the following ansatz for $f_j$ and $f_e$ which have the expected behavior at small $r$, 
\begin{align}
    f_j(r) = -\frac{\alpha_j^2}{r^2}, \qquad f_e(r) = 1 - \frac{ \alpha_e^2}{r^2}+\frac{\tilde{\alpha}_e^4}{r^4} .
\end{align}
We can then try solving the equations of motion for the dilaton, which again gives
\begin{align}
    e^{2\Phi} = c_2^2 \left( \frac{1 }{\alpha_1^2 + k_1 r^2} \right) \, , 
\end{align}
where $c_2$ is an unknown constant, as in the solution (\ref{dilatonsolution}) with $\Jt = 0$. 
%
We also find the algebraic constraints
\begin{align}
    \alpha_x = \sqrt{ \frac{ k_x }{k_1} } \alpha_1 \, , \qquad \tilde{\alpha}_e = \left( \frac{k_1}{k_x} \right)^{1/4} \alpha_j \, , \qquad \alpha_5 = r_5 \, .
\end{align}
As in the discussion around equation \C{final_string_frame_soln}, we can rescale $x_5$ to set $k_x=k_1$ which significantly simplifies the number of parameters. In the spinning case, the relationship between $c_2$ and the other parameters, imposed by the dilaton equation of motion, is modified. In this case we find
\begin{align}
    c_2^4 = \frac{r_5^4 \left( \alpha_1^4 + k_1 \left( \alpha_1^2 \alpha_e^2 + k_1 \alpha_j^4 \right) \right)}{\rh_1^4} \, .
\end{align}
Likewise, the equations of motion determine $f_r ( r )$ to be
\begin{align}
    f_r ( r ) = \frac{r_5^2}{r^2 - \alpha_e^2 + \frac{\alpha_j^4}{r^2}}  \, .
\end{align}
 In determining the parameter map, we see that the relations,
\begin{align}
    \frac{r^2}{\rt^2} = \frac{\a_1^2 r_5^2}{R^2}\, , \qquad  \ell^2 = \frac{\alpha_1^4}{c_2^4} r_5^2\, , 
\end{align}
remain unchanged from the case without spin. With this identification, we can determine $\a_j$ in terms of $\Jt$ appearing in the metric (\ref{BTZJ}). The final string-frame metric and dilaton take the form, 
\begin{align}\label{final_string_frame_soln_J}
   &  ds^2 = - \frac{1 - \frac{ \alpha_e^2}{r^2} + \frac{\alpha_j^4}{r^4}}{k_1 + \frac{\rh_1^2}{r^2}} \, dt^2 + \frac{1}{k_1 + \frac{\rh_1^2}{r^2}} \, \left( dx_5 - \frac{\alpha_j^2}{r^2} dt\right)^2 + \frac{r_5^2}{r^2 - \alpha_e^2 + \frac{\alpha_j^4}{r^2}}  \, dr^2 + r_5^2 \, d \Omega_3^2 + ds^2_{T^4} \, , \nonumber \\
&  e^{2\Phi} = \frac{r_5^2 \sqrt{1 + 8 k_1\frac{ \Mt  r_5^2}{R^2} + 16 \left(k_1\right)^2 \frac{ \Jt^2 r_5^4}{R^4} }}{\rh_1^2 + k_1 r^2} \, , \qquad  \alpha_e^2 = 8 \Mt \frac{\rh_1^2 r_5^2}{R^2}\, , \qquad \a_j^2 = 4 \Jt \, \frac{ \rh_1^2 r_5^2}{R^2} .
\end{align}
The form of the dilaton solution is eerily reminiscent of the solution of the $\TT$-deformed energy flow equation~\C{energyformula}.

\subsection{The asymptotically flat case where $k_5 > 0$}

We will now attempt to recouple the asymptotic region by taking $k_5>0$ for the general class of metrics with mass and spin. This means that asymptotic theory will be $6$-dimensional rather than $3$-dimensional because the $3$-sphere will decompactify as $r\rightarrow \infty$, while the $T^4$ remains compact.

\subsubsection*{\ul{\it Interpolating Ansatz}}

To simplify our analysis, we take the following ansatz for the ten-dimensional metric which is a slight generalization of the prior cases,
\begin{align}\begin{split} \label{metricansatz}
    & f_1(r) = k_1 + \frac{\alpha_1^2}{r^2}, \qquad  f_e(r) = 1 - \frac{ \alpha_e^2}{r^2} + \frac{\tilde{\alpha}_e^4}{r^4}, \qquad f_5(r)  = k_5 + \frac{\alpha_5^2}{r^2},  \\
    & f_x(r) = k_x + {\alpha_x^2 \over r^2}  \, , \qquad  f_j(r) = - \frac{\alpha_j^2}{r^2} \, .
\end{split}\end{align}
We also demand that $k_1>0, k_x>0$ and $k_5>0$ for a sensible metric at large $r$. We can repeat the steps of the preceeding analysis and see what constraints emerge from trying to solve the spacetime equations of motion. Firstly, there are algebraic constraints generalizing what we found in the $k_5=0$ case:
\begin{align}
    \tilde{\alpha}_e = \left( \frac{k_1}{k_x} \right)^{1/4} \alpha_j \, , \qquad \alpha_x = \sqrt{ \frac{ k_x }{k_1} } \alpha_1 \, , \qquad \alpha_e = \frac{ \sqrt{ k_x ( r_5^4 - \alpha_5^4 ) - k_1 k_5^2 \alpha_j^4 }}{\alpha_5 \sqrt{ k_5 k_x } }\, .
\end{align}
Once again it is natural to rescale the $x_5$ circle to set $k_x =k_1$ and replace both $\tilde{\alpha}_e$ by $\a_j$ and $\a_x$ by $\a_1$.  With these replacements the solution for $f_r(r)$ is given by, 
\begin{align}
    f_r ( r ) = \frac{k_5 \alpha_5^2 r^2 ( k_5 r^2 + \alpha_5^2 ) }{ r^2 \left( \alpha_5^4 + k_5 \alpha_5^2 r^2 - r_5^4 \right) + k_5 \alpha_j^4 \left( k_5 r^2 + \alpha_5^2 \right) } \, , 
\end{align}
while the dilaton takes the form:
\begin{align} \label{dilsolution}
     e^{2 \Phi} = \frac{k_5 r^2 + \alpha_5^2}{\rh_1^2 \left( k_1 r^2 + \alpha_1^2 \right)} \cdot \sqrt{ \alpha_1^4 + k_1^2 \alpha_j^4 - \frac{ k_1 \alpha_1^2 \left( \alpha_5^4 + k_5 \alpha_j^4 - r_5^4 \right) }{k_5 \alpha_5^2 } } \, .
\end{align}
It is more convenient to invert the relation between $\a_5$ and $\a_e$,
\begin{align} \label{solalpha5}
    \alpha_5^2 = \frac{1}{2} \left( \sqrt{ k_5^2 \alpha_e^4 + 4 \left( r_5^4  - k_5^2 \alpha_j^4 \right)} - k_5 \alpha_e^2 \right) 
\end{align}
As a check, notice that $\a_5 \rightarrow r_5$ when we take $k_5\rightarrow 0$ which is the solution we found earlier in \C{solveax}. In this more general case, $\a_5$ is mass-dependent. In a small $k_5$ expansion, we see that
\begin{align}
    \a_5^2 = r_5^2 - \frac{k_5 \a_e^2}{2} + O(k_5^2)\, .
\end{align}
Substituting (\ref{solalpha5}) into (\ref{dilsolution}) gives the following nicer form for the dilaton:
\begin{align} \label{nicerdilsolution}
     e^{2 \Phi} = \frac{k_5 r^2 + \alpha_5^2}{k_1 r^2 + \alpha_1^2} \cdot \sqrt{ \frac{\alpha_1^4 + k_1 \alpha_1^2 \alpha_e^2 + k_1^2 \alpha_j^4}{\rh_1^4} } \, .
\end{align}

Let us count parameters versus expectations. First $\a_e^2$ and $\a_j^2$ are determined by the mass and spin of the deep interior BTZ black hole via \C{final_string_frame_soln_J}. Together with $k_5$, these parameters determine $\a_5^2$ using \C{solalpha5}. That leaves $\a_1, k_1$ and $k_5$ to be identified with physical parameters of the asymptotic solution. Now the existence of a non-zero $k_5$ is the statement that we have not taken the partial decoupling limit. Therefore it should be possible to identify $k_5$ with the asymptotic value of the string coupling, called $g_s$ in \C{kluson_soln}. 

To determine the precise identifications, we can start by examining the asymptotic value of the dilaton.  This cannot be mass or spin-dependent because it is part of the data defining the quantum gravity Hilbert space.
Unlike the prior discussion, we now need to make a conformal transformation in a $6$-dimensional gravity theory rather than a $3$-dimensional theory. 
The Einstein frame metric is now given by, 
\begin{align}
   ds^2_{\rm Einstein} =  e^{-(\Phi- \Phi_0)} ds^2_{\rm string}\, ,
\end{align}
where $\Phi_0$ is the now finite $r\rightarrow\infty$ value of the dilaton. We will require that $\Phi_0$ be independent of mass and spin since it determines the gravitational constant. 
To determine how to implement this condition, it is useful to examine the nicer form of the dilaton appearing in \C{nicerdilsolution}. It is not possible to make the square root independent of mass and spin by modifying $\a_1$ in a way proportional to $k_5$ so we expect that the relation
\begin{align}
    \a_1 = \rh_1\, ,
\end{align}
is unchanged. This means we must absorb the mass and spin-dependence in $k_5$. So we define
\begin{align}
    \kh_5 = k_5 \cdot \gamma, \qquad \gamma= \sqrt{ 1 + \frac{k_1 \alpha_e^2}{\alpha_1^2} + \frac{k_1^2 \alpha_j^4}{\alpha_1^4} }\, .
\end{align}
In terms of $\hk_5$, the dilaton now takes the form
\begin{align}
    e^{2 \Phi} = \frac{\kh_5 r^2 + \gamma \alpha_5^2}{k_1 r^2 + \alpha_1^2} \, .
\end{align}
As $r\rightarrow\infty$ the Einstein frame metric now contains the terms, 
\begin{align}
    e^{-(\Phi-\Phi_0)} ds^2_{\rm string}  = \frac{ \kh_5}{\gamma}\left(  dr^2  +  r^2 d\Omega_3^2 \right) + \ldots \, .
\end{align}
The natural radial coordinate at large distances is now
\begin{align}
    \rho^2 = \frac{\kh_5}{\gamma} r^2\, ,
\end{align}
and the dilaton takes the form
\begin{align}
     e^{2 \Phi} &= \frac{\kh_5}{k_1} + \frac{k_1 \alpha_5^2 \gamma - \kh_5 \alpha_1^2}{k_1^2 r^2} + O\left( \frac{1}{r^4} \right) \, , \nonumber \\
     &=  \frac{\kh_5}{k_1} + \frac{\kh_5}{\gamma \rho^2} \frac{k_1 \alpha_5^2 \gamma - \kh_5 \alpha_1^2}{k_1^2} + O\left( \frac{1}{\rho^4} \right)  \, .
\end{align}
The mass-dependence is now encoded in the subleading terms of the $\rho$ expansion. Now we can identify $g_s^2 = \frac{\kh_5}{k_1}$, or equivalently $\kh_5 = k_1 g_s^2$. The periodicity of the $x_5$ circle is $\frac{2\pi R }{\sqrt{k_1}}$ at $r=\infty$. 

The final string-frame metric and dilaton take the form, 
\begin{align}\label{final_undecoupled_soln_J}
   &  ds^2 = - \frac{1 - \frac{ \alpha_e^2}{r^2} + \frac{\alpha_j^4}{r^4}}{k_1 + \frac{\rh_1^2}{r^2}} \, dt^2 + \frac{1}{k_1 + \frac{\rh_1^2}{r^2}} \, \left( dx_5 - \frac{\alpha_j^2}{r^2} dt\right)^2  \nonumber \\
   &\quad  + \frac{k_5 r^2 - \frac{k_5 \alpha_e^2}{2} + \sqrt{ r_5^4 - k_5^2 \alpha_j^4 + \frac{1}{4} k_5^2 \alpha_e^4}}{r^4 - r^2 \alpha_e^2 + \alpha_j^4}  \, r^2  \, dr^2  + r^2 \left( k_5 + \frac{\alpha_5^2}{r^2} \right) \, d \Omega_3^2 + ds^2_{T^4} \, , \nonumber \\
&  e^{2 \Phi} = \frac{g_s^2 + \frac{\gamma \alpha_5^2}{k_1 r^2}}{1 + \frac{\rh_1^2}{k_1 r^2}}  \, , \qquad  \alpha_e^2 = 8 \Mt \frac{\rh_1^2 r_5^2}{R^2}\, , \qquad \a_j^2 = 4 \Jt \, \frac{ \rh_1^2 r_5^2}{R^2} \, , \qquad k_5 = \frac{k_1 g_s^2}{\gamma} \, ,  \nonumber \\
    &\alpha_5^2 = \frac{1}{2} \left( \sqrt{ k_5^2 \alpha_e^4 + 4 \left( r_5^4  - k_5^2 \alpha_j^4 \right)} - k_5 \alpha_e^2 \right)  \, , \qquad \gamma= \sqrt{ 1 + \frac{k_1 \alpha_e^2}{\rh_1^2} + \frac{k_1^2 \alpha_j^4}{\rh_1^4} } \, .
\end{align}
This final form for the metric depends on $(g_s, k_1, \rh_1, r_5)$ along with the mass and spin $(\Mt, \Jt)$ of the interior BTZ black hole.

\newpage
\bibliographystyle{utphys}
\bibliography{master}

\end{document}